\documentclass[10pt,twocolumn,twoside]{IEEEtran}

\usepackage{cite}

\ifCLASSINFOpdf
\else
\fi

\usepackage{cite}
\usepackage{comment} 
\usepackage[cmex10]{amsmath}
\usepackage{amsfonts, amsmath, amsthm, amstext, amssymb, mathrsfs, graphicx, color}
\usepackage{array}
\usepackage{mdwmath}
\usepackage{mdwtab}
\usepackage{algpseudocode}
\usepackage{algorithm}
\usepackage{url}
\usepackage{setspace}
\usepackage{footnote}
\usepackage[skip=0pt]{caption}

\newcommand{\bA}{  \pmb{A}  }
\newcommand{\bB}{  \pmb{B}  }
\newcommand{\bC}{  \pmb{C}  }
\newcommand{\bD}{  \pmb{D}  }
\newcommand{\bE}{  \pmb{E}  }
\newcommand{\bI}{  \pmb{I}  }
\newcommand{\bF}{  \pmb{F}  }
\newcommand{\bG}{  \pmb{G}  }
\newcommand{\bH}{  \pmb{H}  }
\newcommand{\bQ}{  \pmb{Q}  }
\newcommand{\bP}{  \pmb{P}  }
\newcommand{\bS}{  \pmb{S}  }

\newcommand{\bT}{  \pmb{T}  }
\newcommand{\bX}{  \pmb{X}  }
\newcommand{\bY}{  \pmb{Y}  }
\newcommand{\bZ}{  \pmb{Z}  }
\newcommand{\bU}{  \pmb{U}  }

\newcommand{\bW}{  \pmb{W}  }
\newcommand{\bPhi}{  \pmb{\Phi}  }

\newcommand{\bGam}{  \pmb{\Gamma}  }
\newcommand{\bLam}{  \pmb{\Lambda}  }
\newcommand{\bTh}{  \pmb{\Theta}  }

\newcommand{\bth}{  \pmb{\theta}  }
\newcommand{\bpsi}{  \pmb{\psi}  }
\newcommand{\bphi}{  \pmb{\phi}  }
\newcommand{\bgam}{  \pmb{\gamma}  }
\newcommand{\bSig}{  \pmb{\Sigma}  }
\newcommand{\bx}{  \pmb{x}  }
\newcommand{\by}{  \pmb{y}  }
\newcommand{\bw}{  \pmb{w}  }
\newcommand{\bn}{  \pmb{n}  }
\newcommand{\bz}{  \pmb{z}  }
\newcommand{\bp}{  \pmb{p}  }
\newcommand{\bq}{  \pmb{q}  }

\newcommand{\bs}{  \pmb{s}  }
\newcommand{\be}{  \pmb{e}  }
\newcommand{\bg}{  \pmb{g}  }
\newcommand{\ba}{  \pmb{a}  }
\newcommand{\bb}{  \pmb{b}  }
\newcommand{\bff}{  \pmb{f}  }
\newcommand{\br}{  \pmb{r}  }
\newcommand{\bo}{  \pmb{0}  }

\newcommand{\tbGam}{ \tilde{ \pmb{\Gamma} } }
\newcommand{\tbPhi}{ \tilde{ \pmb{\Phi} } }
\newcommand{\tbTh}{ \tilde{ \pmb{\Theta} } }
\newcommand{\tbgam}{ \tilde{ \pmb{\gamma} } }

\newcommand{\tbf}{ \tilde{ \pmb{f} } }

\newcommand{\tbs}{ \tilde{ \pmb{s} } }
\newcommand{\tbw}{ \tilde{ \pmb{w} } }
\newcommand{\tbz}{ \tilde{ \pmb{z} } }

\newcommand{\tbD}{ \tilde{ \pmb{D} } }
\newcommand{\tbE}{ \tilde{ \pmb{E} } }

\newcommand{\tbX}{ \tilde{ \pmb{X} } }
\newcommand{\tbW}{ \tilde{ \pmb{W} } }
\newcommand{\tbY}{ \tilde{ \pmb{Y} } }
\newcommand{\tbT}{ \tilde{ \pmb{T} } }

\newcommand{\snr}{\text{SNR}}
\newcommand{\tr}{ \textup{tr} }
\newcommand{\st}{ \textup{s. t.} }
\newcommand{\argmin}{ \textup{argmin} }
\newcommand{\argmax}{ \textup{argmax} }

\newcommand{\calS}{ \mathcal{S} }
\newcommand{\bbC}{ \mathbb{C} }

\newtheorem{lemma}{Lemma}

\newtheorem{proposition}{Proposition}

\newtheorem{corollary}{Corollary}

\newtheorem{remark}{Remark}

\newcommand{\revised}[1]{\textcolor{black}{#1}}

\newcommand{\ww}[2]{  \pmb{w}^{(#1)}_{#2}  }

\begin{document}

\title{Subspace Estimation and Decomposition for Large Millimeter-Wave MIMO systems}

\author{Hadi~Ghauch,~\IEEEmembership{Student Member,~IEEE,} 
        Taejoon~Kim,~\IEEEmembership{Member,~IEEE,}
        Mats~Bengtsson,~\IEEEmembership{Senior Member,~IEEE,}
        ~and~Mikael~Skoglund,~\IEEEmembership{Senior Member,~IEEE} 
        \thanks{Copyright (c) 2014 IEEE. Personal use of this material is permitted. However, permission to use this material for any other purposes must be obtained from the IEEE by sending a request to pubs-permissions@ieee.org}

		\thanks{H. Ghauch, M. Bengtsson, M. Skoglund, and are with the School of Electrical Engineering and the ACCESS Linnaeus Center, KTH Royal Institute of Technology, Stockholm, Sweden. E-mails: ghauch@kth.se, mats.bengtsson@ee.kth.se, skoglund@kth.se }

		\thanks{T. Kim is with Department of Electronic Engineering, City University of Hong Kong, Kowloon Tong, Hong Kong. E-mail: taejokim@cityu.edu.hk}
    	    
        }

\markboth{}{Ghauch \MakeLowercase{\textit{et al.}}: Subspace Estimation and Decomposition for Large Millimeter-Wave MIMO systems}

\maketitle

\begin{abstract}
Channel estimation and precoding in hybrid analog-digital millimeter-wave (mmWave) MIMO systems  is a fundamental problem that has yet to be addressed, before any of the promised gains can be harnessed. For that matter, we propose a method (based on the well-known Arnoldi iteration) exploiting channel reciprocity in TDD systems and the sparsity of the channel's eigenmodes, to estimate the right  (resp. left) singular subspaces of the channel, at the BS (resp. MS). We first describe the algorithm in the context of conventional MIMO systems, and derive bounds on the estimation error in the presence of distortions at both BS and MS. We later identify obstacles that hinder the application of such an algorithm to the hybrid analog-digital architecture, and address them individually. In view of fulfilling the constraints imposed by the hybrid analog-digital architecture, we further propose an iterative algorithm for subspace decomposition, whereby the above estimated subspaces, are approximated by a cascade of analog and digital precoder/combiner. Finally, we evaluate the performance of our scheme against the perfect CSI, fully digital case (i.e., an equivalent conventional MIMO system), and conclude that similar performance can be achieved, especially at medium-to-high SNR (where the performance gap is less than $5 \%$), however, with a drastically lower number of RF chains ($\sim$ $4$ to $8$ times less). 
\end{abstract}


\IEEEpeerreviewmaketitle

\section{Introduction} \label{sec:intro} 
With the global volume of mobile data expected to increase \emph{by an order of magnitude  between $2013$ and $2019$}, and the volume corresponding to mobile devices outweighing that of all other devices~\cite{Ericsson_mobility}, mobile network operators have the monumental task of meeting this exponentially increasing demand. Given that  spectrum is a scarce and precious resource, future communication systems have to exhibit unparalleled spectral efficiency. Though earlier results date back to~\cite{Ohata_500Mb_00,Ohata_1Gbps_03}, communication systems in the millimeter wave (mmWave) spectrum have been receiving growing interest over the past years. mmWave communication systems have the distinct advantage of exploiting the \emph{huge amounts of unused (and possibly unlicensed) spectrum} in those bands - around $200$ times more than conventional cellular systems. Moreover, the corresponding antennae size and spacing become small enough, such that \emph{tens-to-hundreds of antennas can be fitted on conventional hand-held devices}, thereby enabling  gigabit-per-second communication.

However, the large number of radio frequency (RF) chains required to drive the increasing number of antennas, inevitably incurs a tremendous increase in power consumption (namely by the analog-to-digital converters), as well as added hardware cost. One elegant and promising solution to remedy this inherent problem is to offload part of the precoding/processing to the \emph{analog domain}, via analog precoding (resp.combining), i.e., a network of phase shifters to linearly process the signal at the the base station (BS) (resp. mobile station (MS). This so-called problem of \emph{analog and digital co-design} for beamforming and precoding in low-frequency regime was first investigated in \cite{Zhang_VSP_05, Venkateswaran_analogBF_10}. This architecture was later studied within the context of higher frequency (mmWave) systems in \cite{Ayach_Spatially_14, Alkhateeb_channel_2014, Nsenga_mixedAD_10} - under the name of \emph{hybrid precoding/architecture} - for the precoding problem. A similar setup for the case of beamforming was considered in~\cite{Tsang_Coding_11,  Wang_Beamcodebook_09, Hur_mmWave_13}.

However, several fundamental challenges have to resolved before any of the promised gains can be harnessed, namely, estimating the (large) mmWave channel, and designing the analog/digital precoders and combiners accordingly. We underline the fact that classical training schemes developed for Multiple-input Multiple-output (MIMO) systems are not applicable for that particular case. Moreover, note that our proposed technique encompasses both beamforming and precoding, i.e., it does not depend on the number of streams.

After a series of approximations to the mutual information, and taking into account precoding (excluding the receive combiners),~\cite{Ayach_Spatially_14} derived an optimality condition relating the analog and digital precoders to the optimal unconstrained precoder (i.e., the right singular vectors of the channel), by assuming \emph{full channel state information (CSI) at both the BS and MS}. This assumption was later relaxed in~\cite{Alkhateeb_channel_2014} where an algorithm for estimating the dominant propagation paths was proposed, based on the previously proposed concept of \emph{hierarchical codebooks sounding} in~\cite{Wang_Beamcodebook_09, Hur_mmWave_13}. However, the algorithm requires \emph{a priori knowledge of the number of propagation paths} (i.e. the propagation environment), its \emph{performance is affected by the sparsity level of the channel}, and exhibits relatively elevated complexity.  Finally, it appears rather inefficient to estimate the entire channel, while only a few eigenmodes are needed for transmission: this is particularly relevant in mmWave MIMO channels, since the majority of eigenmodes have negligible power.    

The approach we present here attempts to address the above limitations. The proposed algorithm is based on the well known \emph{Arnoldi Iteration}, exploits channel reciprocity inherent in Time-Division Duplexing (TDD) MIMO systems to gradually build an orthonormal basis for the corresponding Krylov subspace, and \emph{ directly estimates the dominant  left / right singular modes of the channel, rather than the entire channel}. We then propose an iterative method for subspace decomposition, to \emph{ approximate the estimated right (resp. left) singular subspace by a cascade of analog and digital precoder (resp. combiner)}, while taking into account the hardware constraints of this so-called hybrid analog-digital architecture. The subspace estimation (SE) algorithm is based on \emph{BS-initiated echoing}, whereby the BS sends along some beamforming vector, and the MS echoes its received signal back to the BS (using amplify-and-forward), thereby enabling the BS to obtain an estimate of the effective uplink-downlink channel. We first detail the algorithm in the context of conventional MIMO, taking into account distortions in the the system (e.g., noise, or other disturbances),  derive bounds on the estimation error, and highlight its desirable features. We then adapt its structure, to fit the many operational constraints dictated by the hybrid analog-digital architecture. While we feel that aspects such as complexity, overhead and numerical stability are best left for future works, we do shed light on each of them. Although the main results of the paper were earlier presented in~\cite{Ghauch_BSE_conf}, we provide in this work  an in-depth look at our proposed methods, and derive several performance results.

In the following, we use bold upper-case letters to denote matrices, and bold lower-case letters denote vectors. Furthermore, for a given matrix $\pmb{A}$, $[\pmb{A}]_{i:j}$ denotes the matrix formed by taking columns $i$ to $j$, of $\pmb{A}$, $\tr(\pmb{A})$ denotes its trace, $\Vert \pmb{A} \Vert_F^2$ its Frobenius norm, $|\pmb{A}|$ its determinant, $\pmb{A}^\dagger$ its conjugate transpose. $ [\bA]_{i,j} = a_{i,j}$ denotes element $(i,j)$ of $\bA$, $\ba_i$ the $i$th of column $\bA$, and $[\ba]_i = a_i$ element $i$ in vector $\ba$. $ [\bA]_{SL}$ and $[\bA]_{U}$ represent the matrix formed by the strictly lower and upper triangular matrix of a square matrix $\bA$, respectively. $\bI_n$ denotes the $n \times n$ identity matrix, $\textrm{diag}(\bx)$ is a diagonal matrix with elements of $\bx$ on its diagonal, $\Re(x)$ the real part of $x$,  $\sigma_{\max}[\bU]$ / $\sigma_{\min}[\bU]$ the maximum/minimum singular value of $\bU$. Moreover, $ \hat{\bU} = \textrm{qr} (\bU)$ refers to the semi-unitary matrix returned by the QR algorithm, with $\bU^\dagger \bU = \bI$. Finally, we let $\lbrace n \rbrace \triangleq \lbrace 1, ..., n \rbrace $, and $\calS_{p,q} = \left\lbrace \bX \in \bbC^{p \times q} \ | \ \ \vert \bX_{ij} \vert = 1/\sqrt{p} \ , \ \forall (i,k) \in \lbrace p \rbrace \times \lbrace q \rbrace \right\rbrace $.

\section{System Model} \label{sec:sysmod}

\subsection{Signal Model}
\begin{figure}
 \begin{centering}
  \includegraphics[width=9cm,height=2.5cm]{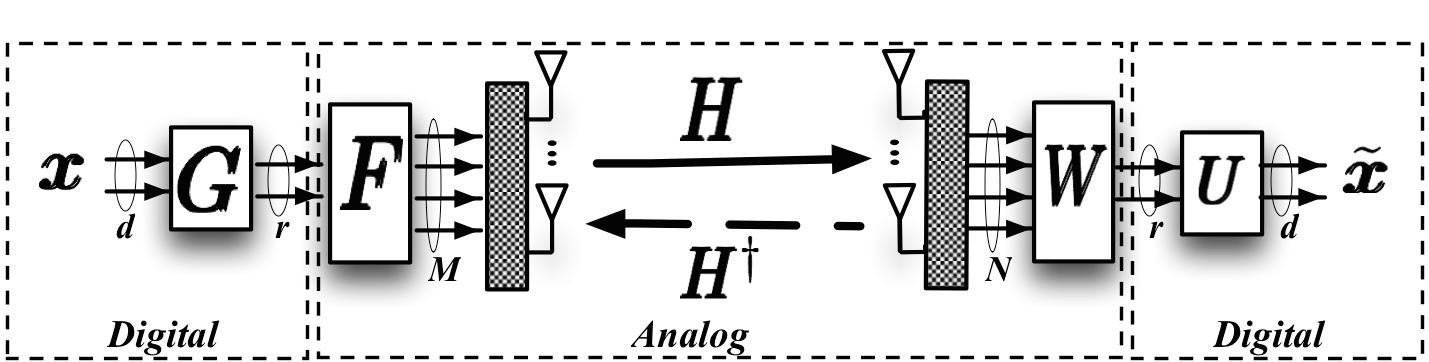}
  \vspace{.2cm}
  \caption{Hybrid Analog-Digital MIMO system architecture} \label{setup}
 \end{centering}
\end{figure}
Assume a single user MIMO system with $M$ and $N$ antennas at the BS and MS, respectively, where each is equipped with $r$ RF chains, and sends $d$ independent data streams (where we assume that  $d \leq r \leq \min(M,N)$). The downlink (DL) received signal is given by 
\begin{align}
\by^{(r)} = \bH \bF \bG \bx^{(t)} + \bn^{(r)}
\end{align}
where  $\bH \in \bbC^{N \times M }$ is the complex channel - assumed to be slowly block-fading,  $\bF \in \mathbb{C}^{M \times r}$ is the analog precoder, $\bG \in \mathbb{C}^{r \times d}$ the digital precoder,  $\by^{(r)}$ the $N$-dimensional signal at the MS antennas, $\bx^{(t)}$ is the $d$-dimensional transmit signal with covariance matrix $E[\bx^{(t)} \bx^{{(t)}^\dagger}] = \bI_d$ and $\bn^{(r)}$ is the AWGN noise at the MS, with $E[\bn^{(r)} \bn^{{(r)}^\dagger}] = \sigma_{(r)}^2 \bI_N $. Note that $^{(t)}$ and $^{(r)}$ subscripts/superscripts denote quantities at the BS and MS, respectively.  Both the analog precoder and combiner are constrained to have constant modulus elements (since the latter represent phase shifters), i.e., $\bF \in \calS_{M,r}$ and $\bW \in \calS_{N,r}$ (also referred to as the \emph{constant-modulus} or \emph{constant-envelope} constraint).  
We adopt a \emph{total power constraint} on the effective precoder, i.e., $\Vert \bF \bG \Vert_F^2 \leq d$, a widespread one in the hybrid analog-digital precoding literature~\cite{Ayach_Spatially_14,Alkhateeb_channel_2014}.
With that in mind, the received signal after filtering in the DL is given as, 
\begin{align}\label{eq:sigmod}
\tilde{\bx} &= \bU^\dagger \bW^\dagger \by^{(r)} = \bU^\dagger \bW^\dagger \bH \bF \bG \bx^{(t)} + \bU^\dagger \bW^\dagger \bn^{(r)} 
\end{align}
where $\bW \in \mathbb{C}^{N \times r}$ and $\bU \in \mathbb{C}^{r \times d}$ are the analog and digital combiners, respectively\footnote{Similarly, exploiting channel reciprocity, the uplink received signal is given by  $\by^{(t)} = \bH^\dagger \bW \bU \bx^{(r)} + \bn^{(t)}$  where $\by^{(t)}$ is the $M$-dimensional signal at the BS and $\bn^{(t)}$ is the AWGN noise at the BS, such that $E[\bn^{(t)} \bn^{{(t)}^\dagger}] = \sigma_{(t)}^2 \bI_M. $}. We also assume a TDD system, where channel reciprocity holds. Finally, we denote the SVD of $\bH$ as,
\begin{align} \label{eq:svdH}
\bH =   
\begin{bmatrix}
\bPhi_1 , & \bPhi_2
\end{bmatrix}
\begin{bmatrix}
\bSig_1  & \bo \\
\bo & \bSig_2
\end{bmatrix}
\begin{bmatrix}
\bGam_1^\dagger \\
\bGam_2^\dagger 
\end{bmatrix}
= \bPhi_1 \bSig_1 \bGam_1^\dagger + \bPhi_2 \bSig_2 \bGam_2^\dagger
\end{align}
where $\bGam_1 \in \bbC^{M \times d}$ and $\bPhi_1 \in \bbC^{N \times d} $ are semi-unitary, and $\bSig_1 = \textrm{diag}(\sigma_1, ..., \sigma_d) $ is diagonal with the $d$ largest singular values of $\bH$ (in decreasing order).

\subsection{Motivation}
Keeping in line with previous work in that area, our aim is to design the precoders and combiners as follows,
\begin{align} \label{opt:min_euc_dist}
	(\bF^\star , \bG^\star ) &= 
             \begin{cases}
                 \underset{\bF  , \ \bG  }{\min} \ \  \Vert \bGam_1 - \bF \bG  \Vert_F^2   \\
                 \st \  \Vert \bF \bG \Vert_F^2 \leq d , \ \ \bF \in \calS_{M,d}                 
              \end{cases} \nonumber \\
	(\bW^\star , \bU^\star ) &= 
               \begin{cases}
                 \underset{\bW  , \ \bU  }{\min} \  \Vert \bPhi_1 - \bW \bU  \Vert_F^2   \\
                 \st \ \ \ \bW \in \calS_{N,d}                 
              \end{cases}
\end{align}
The latter design criterion has been quite prevalent in earlier works relating to the hybrid analog-digital architecture, and applied rather successfully in~\cite{Ayach_lowcomplexity_12,Ayach_Spatially_14,Nuria_hybrid_15,Alkhateeb_channel_2014}. After a series of approximations to the mutual information in~\cite{Ayach_Spatially_14}, it was shown that the optimal precoders, $\bF, \bG$, are formulated in exactly the same fashion as above (though their formulation did not include receive combining).

Moreover, we use the following expression as a performance metric (i.e., the ``user-rate'' corresponding to a given choice of precoders and combiners), 
\begin{align} \label{eq:rate}
R &= \log_2 \left| \bI_d + \bH_{e} \bH_{e}^\dagger   ( \sigma_{(r)}^2 \bU^\dagger \bW^\dagger \bW \bU)^{-1}\right| 
\end{align}
where $\bH_{e} = \bU^\dagger \bW^\dagger \bH \bF \bG$, $\frac{1}{\sigma_{(r)}^2} \triangleq \textrm{SNR}$ is the signal-to-noise ratio. Moreover we assume, for simplicity, that uniform power allocation is performed (no waterfilling), keeping in mind that a power allocation matrix $\bLam$ can be easily incorporated in the expression.  
Although not directly optimized, the above expression was used in~\cite{Ayach_Spatially_14}, within the context of hybrid analog-digital precoding. 
As we will discuss below, the value of the expression in~\eqref{eq:rate} is related to  achievable rates over the considered hybrid analog-digital MIMO link; in particular $R$ becomes an \emph{achievable rate} in the scenario that both the BS and MS are provided perfect knowledge of $\bH$.

In a nutshell,~\eqref{opt:min_euc_dist} boils down to finding $\bF\bG$ (resp. $\bW \bU$) that ``best'' approximate $\bGam_1$ (resp. $\bPhi_1$). Moreover, if there exists optimal precoders and combiners that make the distances in~\eqref{opt:min_euc_dist} zero, then they must satisfy  $$ \bF^\star \bG^\star  = \bGam_1 , \ \ \ \ \bW^\star \bU^\star  = \bPhi_1. $$ We denote by $R^\star$ the resulting ``user-rate'' that is obtained by plugging in the above precoders/combiners in~\eqref{eq:rate}. Then $R^\star$ can be expressed as,
\begin{align} \label{eq:optrate}
 R^\star \triangleq R(\bF^\star , \bG^\star , \bW^\star , \bU^\star) =  \log_2 \left| \bI_d + \snr \ \bSig_1^2 \right|
\end{align} 
Following the above discussion on the achievability of $R$, $R^\star$ is the \emph{maximum achievable rate} over the precoders and combiners, when $\bH$ is known to both BS and MS.
We underline the fact that $R$ in~\eqref{eq:rate} depends on the subspace spanned by the precoders / combiners, rather than the Euclidean distance between the right/left dominant subspace and the  precoder/combiner, i.e.,~\eqref{opt:min_euc_dist}.
However, optimizing metrics that involve span or chordal distances, is not straightforward.  We thus  emphasize that attempts at directly maximizing $R$ in~\eqref{eq:rate} are outside the scope of this work: rather, the focus is put on proposing  mechanisms for subspace estimation and decomposition, and analyzing their performance.

Moreover, since we assume that no channel information is available at neither the BS, nor the MS, our aim is \emph{firstly to obtain an estimate of the subspaces in question}, i.e. $\tbPhi_1 \approx \bPhi_1$ at the MS, and $\tbGam_1 \approx \bGam_1 $ at the BS. We then propose methods that optimize the precoders and combiners to \emph{accurately approximate the estimated subspaces}, by providing means to solve problems such as $\Vert \tbGam_1 - \bF \bG \Vert_F^2$ and $\Vert \tbPhi_1 - \bW \bU \Vert_F^2$ (while taking into consideration the constraints inherent to the hybrid analog-digital architecture).

\section{Eigenvalue Algorithms and Subspace Estimation} \label{sec:bsemimo}
\subsection{Subspace Estimation vs. Channel Estimation} \label{sec:motiv}
The aim of subspace estimation (SE) methods in MIMO systems is to estimate a predetermined \emph{low-dimensional subspace of the channel}, required for transmission. We illustrate this in the context of conventional MIMO systems, i.e., where precoders/combiners are fully digital. For the sake of exposition, we start with a simple toy example, where noiseless single-stream transmission is assumed (and ignoring any physical constraints). The BS selects a random unit-norm beamforming vector, $\bp_1$, and then sends $\bp_1 x^{(t)} $, where $x^{(t)}=1$.  The received signal, $\bq_1 = \bH \bp_1$, is echoed back to the BS (in effect, this implies that the signal is complex conjugated before being sent), in an Amplify-and-Forward (A-F) like fashion.\footnote{This mechanism for MIMO subspace estimation, where the MS echoes back the transmitted signal using A-F, was first reported in~\cite{Dahl_blind_04}.} Then, exploiting channel reciprocity, the received signal at the BS is first normalized, i.e., $\bp_2 = \bH^\dagger \bq_1/\Vert \bH^\dagger \bq_1 \Vert_2 =  \bH^\dagger \bH \bp_1 / \Vert \bH^\dagger \bH \bp_1 \Vert_2 $,  and then echoed back to the MS. This simple procedure is done iteratively, and the resulting  sequences $\lbrace \bp_l \rbrace$  at the BS, and $\lbrace \bq_l \rbrace$ at the MS, are defined as follows,
\begin{align} \label{eq:pm}
\bp_{l+1} &= \bH^\dagger \bH \bp_l / \Vert \bH^\dagger \bH \bp_l \Vert_2 ; \ \ 
\bq_{l+1} = \bH \bp_l
\end{align}
It was noted in~\cite{Dahl_blind_04} that using the Power Method (PM), one can show that as $l \rightarrow \infty$, $\bp_l \rightarrow \bgam_1$ and $\bq_l \rightarrow \sigma_1 \bphi_1 $, implying that this seemingly simple ``ad-hoc'' procedure will converge to the \emph{maximum eigenmode transmission}. The authors of~\cite{Dahl_blind_04} also generalized the latter method to multistream transmission, i.e., by estimating $\bGam_1$ and $\bPhi_1$, using the Orthogonal/Subspace Iteration (which was dubbed Two-way QR (TQR) in~\cite{Dahl_blind_04, Dahl_intrinsic_07}).  


We note that SE schemes such as the ones described above, offer the following distinct advantage over classical \emph{pilot-based channel estimation}:  in spite of the large number of transmit and receive antennas, SE methods can estimate the dominant left/right singular subspaces with a relatively low communication overhead, when the latter have small dimension (relative to the channel dimensions). Consequently, subspace estimation is much more efficient than channel estimation, especially in large low-rank MIMO systems such as mmWave channels (because the latter estimates the dominant low-dimensional subspace instead of the whole channel). For the reason above, our proposed algorithm falls under the umbrella of SE methods. We first describe this algorithm in the context of ``classical'' MIMO systems, and later adapt it to the hybrid analog-digital architecture.

\subsection{Arnoldi Iteration for Subspace Estimation} \label{sec:mimobse}
Despite the fact that Krylov subspace methods (such as the Arnoldi and Lanczos Iterations for symmetric matrices) are among the most common methods for eigenvalue problems~\cite{Watkins_eigenvalue_07}, their use in the area of channel/subspace estimation is limited to  equalization for doubly selective OFDM channels~\cite{Hrycak_equalization_10}, and channel estimation in CDMA systems~\cite{Torlak_kyrlov_02}. Algorithms falling into that category iteratively build a \emph{basis for the Krylov subspace}, $\mathcal{K}^m = span \lbrace \bx, \bA \bx, ..., \bA^{m-1} \bx  \rbrace $, one vector at a time. We use one of many variants of the so-called \emph{Arnoldi Iteration/Procedure}, and a simplified version of the latter is shown in Table~\ref{alg:arn} (as presented in~\cite{Saad_Numerical_11}). 
\begin{table} 
\begin{algorithmic} 
\State Set $m$ ($m \leq M$); $\bq_1 = $ random unit-norm ; $\bQ = [\bq_1]$
\For {$l = 1, 2, ..., m $}
	\State  1.a $\bp_l = \bA \bq_l$
	\State  1.b $t_{k,l} = \bq_k^\dagger  \bp_l , \ k = 1, \dots, l $
	\State  2. $\ \br_l = \bp_l - \sum_{k=1}^l t_{k,l}\bq_k $ 
	\State  3. $\ t_{l+1, l} = \Vert \br_l \Vert_2 $ ; \textbf{if} $(t_{l+1, l} = 0)$ \textbf{stop}
	\State  4. $\ \bQ = [ \bQ ,  \  \bq_{l+1} = \br_l/t_{l+1, l} ]$
\EndFor
\end{algorithmic}
\vspace{.2cm}
\caption{Arnoldi Procedure}  \label{alg:arn}
\end{table}
The algorithm returns $\bQ_m = [\bq_1 , \dots,  \bq_m ] \in \bbC^{M \times m} $ and an upper Hessenberg matrix $\bT_m \in \bbC^{m \times m} $, such that  $$\bQ_m^\dagger \bA \bQ_m = \bT_m , \ \bQ_m^\dagger \bQ_m = \bI_m. $$
It can be shown that the algorithm iteratively builds $\bQ_m$, an orthonormal basis for $\mathcal{K}^m$ (when roundoff errors are neglected), and that $\bQ_m^\dagger \bA \bQ_m = \bT_m$. 
We then say that the eigenvalues/eigenvectors of $\bT_m$ are called \emph{Ritz eigenvalues/eigenvectors}, and approximate the eigenvalues/eigenvectors of $\bA$. The main idea behind processes such as the Arnoldi (and Lanczos) is to find the dominant eigenpairs of $\bA$, by finding the eigenpairs of $\bT_m$. 

We note that the Arnoldi algorithm is a generalization of the Lanczos algorithm for the non-symmetric case, i.e., the latter is specifically tailored for cases where $\bA \succeq \pmb{0} $ (this is clearly the case in this work, since $\bA = \bH^\dagger \bH $ ). This being said, the reason for not using the Lanczos iteration is that in practice, noise that is inherent to the echoing process, makes the Lanczos algorithm not applicable: namely, the requirement that $\bT_m$ is tridiagonal, is violated.

Our goal in this section is to first apply the above algorithm to estimate the $d$ largest eigenvectors of $\bA = \bH^\dagger \bH$ at the BS (which are exactly $\bGam_1$), by implementing a \emph{distributed version of the Arnoldi process}, that exploits the channel reciprocity inherent to TDD systems. Moreover, we extend the original formulation of the algorithm to incorporate a \emph{distortion variable} (representing noise, or other distortions, as will be done later). 

It becomes clear at this stage, that the BS requires knowledge of the sequence $\lbrace\bH^\dagger \bH \bq_l \rbrace_{l=1}^m$, needed for the matrix-vector product in step 1 (Table~\ref{alg:arn}): the latter can be accomplished by obtaining an estimate $\bp_l$, of $\bH^\dagger \bH \bq_l, \ l \in \lbrace m \rbrace$. Without any explicit CSI at  neither the BS nor the MS, we exploit the reciprocity of the medium to obtain such an estimate, via  \emph{BS-initiated echoing}: the BS sends $\bq_l$ over the DL channel, the MS echoes back the received signal in an A-F like fashion, over the uplink (UL) channel (following the process proposed in~\cite{Withers_echomimo_08}, and detailed in Sect.~\ref{sec:motiv}), i.e.,
\begin{align} \label{eq:echo}
DL: \ \ \bs_l &= \bH \bq_l + \ww{r}{l} \nonumber   \\
UL: \ \ \bp_l &= \bH^\dagger \bs_l + \ww{t}{l} = \bH^\dagger \bH \bq_l + \bH^\dagger \ww{r}{l} + \ww{t}{l} \nonumber  \\
&= \bH^\dagger \bH \bq_l + \tbw_l
\end{align}   
where $\bs_l$ is the received signal in the DL, $\ww{t}{l}$ and $\ww{r}{l}$ are distortions at the BS and MS, respectively (representing noise for example). 

After the echoing phase, the BS has an estimate, $\bp_l$, of $\bH^\dagger \bH \bq_l$, as seen from~\eqref{eq:echo}. The remainder of the algorithm follows the conventional Arnoldi Iteration, and is shown in the Subspace Estimation using Arnoldi (SE-ARN) procedure (Table~\ref{alg:distarn}). In addition to $\bT_m$ at the output of the algorithm, we define the matrices, $\tbT_m$, $\tbW_m$ and $\tbE_m$, as follows, 
\begin{align} \label{eq:defs}
[\tbT_m]_{i,l} &= \begin{cases}
               \bq_i^\dagger \bH^\dagger \bH \bq_l, \ \textrm{if} \ l \leq m , \ \forall i \leq l   \\
               \Vert \br_l \Vert_2,  \ \ \textrm{if} \ l < m , \ i = l+1   \\
               0, \ \textrm{otherwise}
            \end{cases} \nonumber \\
\tbW_m &= [\tbw_1, ..., \tbw_m]  , \ \tbE_m = [\bQ_m^\dagger \tbW_m]_{SL}      
\end{align} 
where  $\br_l$ is given in Step 2.b (Table~\ref{alg:distarn}). Note that similarly to the conventional Arnoldi Iteration, $\tbT_m$ is an the upper Hessenberg matrix. It then follows from the above definitions that 
\begin{align} \label{eq:bTm_vs_tbTm}
\bT_m = \tbT_m + [\bQ_m^\dagger \tbW_m ]_U .
\end{align}
 This can be easily verified by plugging in Step 1.b into 2.a in Table~\ref{alg:distarn}.  

\begin{table}  
\begin{algorithmic}
\Procedure{$\tbGam_1, \tilde{\bSig}_1 =$ SE-ARN }{$\bH$, $d$}  
\State Set $m$ ($m \leq M$);  Random unit-norm $\bq$;  $\bQ = [\bq_1]$
\For {$l = 1, 2, ..., m $}
	\State // \emph{BS-initiated echoing: estimate $\bH^\dagger \bH \bq_l$}
	\State 1.a \hspace{.1cm} $\bs_l = \bH \bq_l + \ww{r}{l}  $
	\State 1.b \hspace{.1cm} $\bp_l = \bH^\dagger \bs_l + \ww{t}{l} $
	\State // \emph{Gram-Schmidt orthogonalization}
	\State 2.a \hspace{.1cm} $t_{k,l} = \bq_k^\dagger \bp_l \hspace{.7cm}  , \forall \ k = 1, \dots, l $
	\State 2.b \hspace{.1cm} $\br_l = \bp_l - \sum_{k=1}^l \bq_k t_{k,l} $
	\State 2.c \hspace{.1cm} $t_{l+1,l} = \Vert \br_l \Vert_2 $
	\State // \emph{Update $\bQ$} 
	\State 3.a \hspace{.1cm} $\bQ = [ \bQ ,  \  \bq_{l+1} = \br_l/t_{l+1, l} ]$
\EndFor
\State // \emph{Compute $\tbGam_1$} 
\State  \hspace{.1cm} $ \bT_m = \tbTh \tilde{\bLam} \tbTh^{-1}$ 
\State  \hspace{.1cm} $\tilde{\bGam}_1 = \textrm{qr} (\bQ_m \tilde{\bTh}_{1:d})$
\State  \hspace{.1cm} $[\tilde{\bSig}_1]_{i,i} = \sqrt{|  [\tilde{\bLam}]_{i,i}  |} , \forall \ i  $
\EndProcedure
\end{algorithmic}
\vspace{.2cm}
\caption{Subspace Estimation using Arnoldi Iteration (SE-ARN)}   \label{alg:distarn}
\end{table}

At the output of the SE-ARN procedure, the dominant eigenpairs of $\bH^\dagger \bH $ are approximated by those of $\bT_m$ as follows. Let $ \bT_m = \tbTh \tilde{\bLam} \tbTh^{-1}$ be eigenvalue decomposition of $\bT_m$, where $\tilde{\bTh}$ is the (possibly non-orthonormal) set of eigenvectors. Then, it can easily be shown that $\tbGam_1 = \textrm{qr}( \bQ_m [\tbTh]_{1:d})$ are the Ritz eigenvectors of $\bH^\dagger \bH$, where  $[\tbTh]_{1:d}$ has as columns the eigenvectors of $\bT_m$ associated with the $d$ largest eigenvalues (in magnitude).\footnote{Note that, to be exact, the Ritz eigenvectors do not contain any estimation noise. That being said, we stick to this nomenclature, with a slight abuse of definition.
Moreover, $\tilde{\bSig}_1$, the Ritz eigenvalues of $\bH^\dagger \bH$, come for free once the Ritz eigenvectors are obtained (Table~\ref{alg:distarn}).}
 Note that the latter procedure results in the BS obtaining $\tbGam_1$, and consequently $\tilde{\bSig}_1$, using the so-called BS-initiated echoing. This same procedure can be applied using MS-initiated echoing, to estimate $\tbPhi_1$ (i.e., the eigenvectors of $\bH \bH^\dagger$), at the MS.

\subsection{Perturbation Analysis}
In what follows, we extend some of the known properties of the conventional Arnoldi iteration, to account for the estimation error, emanating from the distortion variable. 
\begin{lemma} \label{lem:arnlem}
For the output of the Arnoldi process the following holds,

\noindent$(P1):$ 
\begin{align}
\bQ_m^\dagger \bA \bQ_m = \tbT_m - \tbE_m \triangleq \bC_m, \label{eq:p1}
\end{align}
where $\bC_m = \bS_m \bLam_m \bS_m^{-1} $ is such that $[\bLam]_{i,i} \geq 0$ and $\bS_m^{-1} = \bS_m^\dagger $ \\

\noindent$(P2):$ Let $(\lambda_i^{(m)} , \bs_i^{(m)})$ be any eigenpair of $\bC_m$. Then $(\lambda_i^{(m)} , \ \ \bth_i^{(m)} \triangleq \bQ_m \bs_i^{(m)})$ is an approximate Ritz eigenpair for $\bA$. Furthermore, the approximation error is such that,  
\begin{align}
 \Vert  \bA \bth_i^{(m)} - \lambda_i^{(m)} \bth_i^{(m)} \Vert_2^2 \leq c_m^{(i)} + \Vert \bI_M - \bQ_m \bQ_m^\dagger \Vert_F^2 \Vert \tbW_m \Vert_F^2, \label{eq:p2}
\end{align}
where $c_m^{(i)}= ([\tbT_m]_{m+1,m} | [\bs_i^{(m)}]_m | )^2 $. \\

\noindent$(P3):$ As $ m \rightarrow M$, $\Vert  \bA \bth_i^{(m)} - \lambda_i^{(m)} \bth_i^{(m)} \Vert_2^2 \rightarrow 0$, implying that the  eigenpairs of $\bC_m$ perfectly approximate the eigenpairs of $\bA$(up to round-off errors).  
\end{lemma}
\begin{IEEEproof} 
The proof is shown in Appendix \ref{app:1}. 
\end{IEEEproof} 

We underline the fact that if the distortion variable $\tbW_m$ is zero, the above derivations reduce to the well-known results on the Arnoldi process~\cite[Sect.~6.2]{Saad_Numerical_11}. Lemma~\ref{lem:arnlem} establishes the fact that each eigenpair $(\lambda_i^{(m)} , \bs_i^{(m)})$ of $\bC_m$, is associated with one eigenpair $(\lambda_i^{(m)} , \bth_i^{(m)})$ of $\bA$.\footnote{Though $(P3)$ in Lemma~\ref{lem:arnlem} implies that the error in approximating the eigenpairs of $\bA$ with those of $\bC_m$ vanishes as $m \rightarrow M$, our simulations will later show that very good approximations can be obtained, even for $m \ll M$.}

Thus, one might be tempted to conclude at this point, that by computing the eigenpairs of $\bC_m$, one can \emph{perfectly estimate} the eigenpairs of $\bA$, despite the presence of the distortion variable $\tbW_m$. However, the fact remains that \emph{$\bC_m \triangleq \tbT_m - \tbE_m$ cannot be computed}, mainly because $\tbE_m$ is not known to the BS. As a result,\emph{ $\bT_m$ at the output of the Arnoldi process will be used instead to approximate the eigenpairs of $\bA$}. Now that we established that the eigenpairs of $\bC_m$ approximate that of $\bA$, the natural question is \emph{how close are the eigenpairs of $\bT_m$, to that of $\bC_m$}.

For that purpose, we first show the following,
\begin{align}
\bC_m + \bQ_m^\dagger \tbW_m &= (\tbT_m - \tbE_m ) + \bQ_m^\dagger \tbW_m \nonumber \\
&= \tbT_m  + (\bQ_m^\dagger \tbW_m  - [\bQ_m^\dagger \tbW_m]_{SL} )   \nonumber \\
& = \tbT_m  + [\bQ_m^\dagger \tbW_m]_U  \triangleq \bT_m 
\end{align}
where the first equality follows from the definition of $\bC_m$, and the last one from~\eqref{eq:bTm_vs_tbTm}. 
Thus $\bC_m$ can be viewed as the matrix in question, and $\bP_m \triangleq \bQ_m^\dagger \tbW_m $ a perturbation matrix. We then apply the Bauer-Fike Theorem~\cite[Th.~7.2.2]{golub_matcomp_96} to bound the difference in eigenvalues.

\begin{lemma} \label{lem:pert}
Every eigenvalue $\tilde{\lambda}$ of $\bT_m = \bC_m + \bP_m $ satisfies  $$\vert \tilde{\lambda} - \lambda \vert  \leq \sqrt{m} \ \Vert \tbW_m \Vert_F, $$ where $\lambda$ is an eigenvalue of $\bC_m$. 
\end{lemma}
\begin{IEEEproof}
Refer to Appendix~\ref{app:2} 
\end{IEEEproof}
Summarizing thus far, Lemma~\ref{lem:arnlem} showed that the eigenpairs of $\bA$ can be approximated by the eigenvalues of $\bC_m$, with arbitrarily small error. However, since the latter is not available, we approximate the eigenpairs of $\bC_m$ (and consequently of $\bA$) by those of $\bT_m$, the upper Hessenberg matrix at the output of the Arnoldi process. Finally, Lemma~\ref{lem:pert} established the fact that this approximation error, for the eigenvalues, is upper bounded by the magnitude of the perturbation itself.
We note that the relevant ``error-metric'' here is the distance between the true subspace  $\bGam_1$, and  estimated subspace $\tilde{\bGam}_1  \propto \bQ_m \tilde{\bTh}_{1:d}$ (Table~\ref{alg:distarn}). This does suggest that the estimation error is dependent on $ \tilde{\bTh}_{1:d}$, the eigenvectors of $\bT_m$.  However, performing a similar sensitivity analysis on the eigenvectors is much more involved, since the sensitivity of eigenvectors generally depends on the clustering of eigenvalues.

\section{Hybrid Analog-Digital Precoding for mmWaveMIMO systems} 
In this section we turn our attention to applying the above framework for subspace estimation and precoding, to the hybrid analog-digital architecture. As this section will gradually reveal, several obstacles have to be overcome for that matter. We start by presenting some preliminaries that will be used throughout this section. 

\subsection{Preliminaries: Subspace Decomposition} \label{sec:dcmp}
We will limit our discussion to the digital and analog precoder, keeping in mind that the same applies to the digital and analog combiner. In conventional MIMO systems, the estimates of the right and left singular subspace, $\tbGam_1$ and $\tbPhi_1$, obtained using SE-ARN, can directly be used to diagonalize the channel. However, the hybrid analog-digital architecture entails a cascade of analog and  digital precoder. Thus, $\tbGam_1$ has to be decomposed  into $\bF \bG$ (hence the term \emph{Subspace Decomposition (SD)}), as follows,
\begin{align} \label{opt:qp}
  \begin{cases}
               \underset{\bF, \ \bG}{\min} \ \ h_0(\bF, \bG) = \Vert \tbGam_1 - \bF \bG  \Vert_F^2   \\
               \st \ \ h_1(\bF, \bG) = \Vert \bF \bG \Vert_F^2 \leq d \\
               \hspace{.8cm} \bF \in \calS_{M,d}
            \end{cases}
\end{align}
We underline the fact that the authors in~\cite{Ayach_Spatially_14} arrived to the same formulation as~\eqref{opt:qp}, and proposed a variation on the well-known Orthogonal Matching Pursuit (OMP), to tackle it. The same framework was recently extended in~\cite{Nuria_hybrid_15} to relax the need for dictionaries based on the array response matrix. An alternate decomposition was proposed by~\cite{Yu_hybrid_15}, where the optimization metric is the user rate. Both works were published after the initial submission of our paper.

Within the context of hybrid precoding, the authors in~\cite{Zhang_VSP_05} showed that there exists (non-unique) $\bF \in \calS_{M,r} , \bg \in \bbC^{r \times 1} $ such that $\tbGam_1 = \bF \bg, $ if and only if $r \geq 2$. This was extended in~\cite{Nuria_hybrid_15} where it was shown that there exists $\bF \in \calS_{M,r} , \bG \in \bbC^{r \times d} $ such that $\tbGam_1 = \bF \bG, $ if $r \geq 2d$. We note that for such cases, the cost function in~\eqref{opt:qp} is zero, and we refer to such cases as \emph{optimal decomposition} -whose performance we evaluate in the numerical results section: although the aforementioned schemes use all the available RF chains for the decomposition (and our decomposition uses a subset of the RF chains), the sum-rate performance is actually the same. 

To a certain extent, \eqref{opt:qp} is reminiscent of formulations arising from areas such as blind source separation, (sparse) dictionary learning, and vector quantization ~\cite{Xu_bcd_13, Aharon_ksvd_06}. Though there is a battery of algorithms and techniques that have been developed to tackle such problems, the additional hardware constraint on $\bF$, i.e. $\bF \in \calS_{M,r}$ makes the use of such tools not possible. As a result, we will resort to developing our own algorithm. In spite of the non-convex and non-separable nature of the above quadratically-constrained quadratic program, we propose an iterative  method that attempts to determine an approximate solution.

\subsubsection{Block Coordinate Descent for Subspace Decomposition} \label{sec:bcdsd}
In this part, we further assume that only $d$ of the $r$ available RF chains are used, i.e., $\bF \in \bbC^{M \times d}$ and $\bG \in \bbC^{d \times d}$ (the reason for that will become clear later in this section). The coupled nature of the objective and constraints in~\eqref{opt:qp} suggests a Block Coordinate Descent (BCD) approach. The main challenges arise from the coupled nature of the variables in the constraint (since the latter makes convergence claims of BCD, not possible~\cite{Razaviyayn_BCD_12}), and from the hardware constraint on $\bF$. We will show that a BCD approach implicitly enforces the power constraint in~\eqref{opt:qp}, and consequently the latter can be dropped without changing the problem. 

Our approach consists in relaxing the hardware constraint on $\bF$, and then applying a Block Coordinate Descent (BCD) approach to alternately optimize $\bF$ and $\bG$ (while projecting each of the obtained solutions for $\bF$ on $\calS$). For that matter, we first define the \emph{Euclidean  projection} on the set $\calS$ in the following proposition. 
\begin{proposition} \label{prop:1}
Let $\bX \in \bbC^{M \times d}$ be defined as $[\bX]_{i,k} = |x_{i,k}| \ e^{j\phi_{i,k}}, \ \forall (i,k) $, and $$\bY = \Pi_{\calS}[\bX] \overset{\triangle}{=} \ \underset{\bU \in \calS_{M,d} }{\argmin} \ \Vert \bU - \bX \Vert_F^2  $$ denote its (unique) Euclidean projection on the set $\calS_{M,d}$.  Then $[\bY]_{i,k} = (1/\sqrt{M})\ e^{j\phi_{i,k}},  \forall (i,k) $.
\end{proposition}
\begin{IEEEproof} The proof is straightforward variation on previous results such as~\cite{Zhang_VSP_05}.  \end{IEEEproof}
The latter result implies that given an arbitrary $\bF$, finding the closest point to $\bF$, lying in $\calS_{M,d}$ simply reduces to \emph{setting the magnitude of each element in $\bF$, to $1/\sqrt{M}$}.

Neglecting the constraint on $\bF$ in~\eqref{opt:qp}, one can indeed show that for fixed $\bG$ (resp. $\bF$), the resulting subproblem is convex in $\bF$ (resp. $\bG$). With this in mind, our aim is to produce a \emph{sequence of updates, $\lbrace \bF_k, \bG_k \rbrace_k$ such that the sequence $\lbrace h_0(\bF_k , \bG_k ) \rbrace_k $ is non-increasing} (keeping in mind that monotonicity cannot be shown due to the coupling in the power constraint). Thus, given $\bG_{k}$, each of the updates, $\bF_{k+1}$ and $\bG_{k+1}$, are defined as  as follows,
\begin{align*}
&(J1) \ \ \bF_{k+1} \triangleq \underset{\bF}{\min} \ \ h_0(\bF) = \Vert \tbGam_1 - \bF \bG_k  \Vert_F^2   \\
&(J2) \  \ \bG_{k+1} \triangleq \underset{\bG}{\min} \ \ h_0(\bG) = \Vert \tbGam_1 - \bF_{k+1} \bG  \Vert_F^2
\end{align*}
Both $(J1)$ and $(J2)$ are instances of a non-homogeneous (unconstrained) convex quadratically-constrained quadratic programming (QCQP) that can easily be solved (globally) by finding stationary points of their respective cost functions, to yield,
\begin{align} \label{eq:Fopt}
&\bF_{k+1} =  \tbGam_1 \bG_k^\dagger (\bG_k \bG_k^\dagger)^{-1} 
\end{align}
\begin{align} \label{eq:Gopt}
&\bG_{k+1} = (\bF_{k+1}^\dagger \bF_{k+1})^{-1} \bF_{k+1}^\dagger \tbGam_1 
\end{align}
We note that our earlier assumption that only $d$ of the RF chains are used here (i.e. $\bG$ is square), guarantees that, $(\bG_l \bG_l^\dagger)$ in~\eqref{eq:Gopt} is invertible, almost surely: in fact, our numerical results show that the incurred performance loss is quite negligible. 

Moreover, note that the solution in~\eqref{eq:Fopt} does not necessarily satisfy the hardware constraint on $\bF$. Thus, the result of Proposition~\ref{prop:1} can be used to compute the projection of $\bF$ on $\calS_{M,d}$. To prove our earlier observation that the optimal updates $\bF_{k+1}$ and $\bG_{k+1}$ satisfy the power constraint in~\eqref{opt:qp}, we plug~\eqref{eq:Gopt} into the following (dropping all subscripts for simplicity), 
\begin{align} \label{eq:pc}
\Vert \bF \bG \Vert_F^2 &= \tr \left( \tbGam_1^\dagger \bF \underbrace{ (\bF^\dagger \bF)^{-1} \bF^\dagger \bF }_{=\bI_d} (\bF^\dagger \bF)^{-1} \bF^\dagger \tbGam_1 \right) \nonumber \\
&\leq \tr \left( (\bF^\dagger \bF)^{-1} \bF^\dagger \bF \right) \tr \left( \tbGam_1 \tbGam_1^\dagger \right) = d
\end{align}
where we assumed that $\Vert \tbGam_1 \Vert_F^2 = 1$ w.l.o.g., and used the fact that $\tr(\bA \bB) \leq \tr(\bA)\tr(\bB)$ for $\bA, \bB \succeq \pmb{0} $.
Note that the above relation holds for any arbitrary full-rank $\bF$, and thus, the power constraint is satisfied even after applying the projection step.
The above shows that if  BCD is used, then the power constraint in~\eqref{opt:qp} is always enforced. The corresponding method is termed Block Coordinate Descent for Subspace Decomposition (BCD-SD), and is shown in Table~\ref{alg:bcd_dcmp}. 
\begin{remark} \rm
 We underline the fact that due to the projection step, one cannot show that the sequence $\lbrace h_o( \bF_{k}, \bG_k ) \rbrace_k $ is non-increasing. Nevertheless, despite the fact that monotonic  convergence of BCD-SD cannot be showed analytically, our simulations  indicate that the latter is indeed the case, under normal operating conditions. 
\end{remark}

\begin{remark} \rm
It can be easily verified that the optimal $\bF^\star , \bG^\star$ that maximize $R$ in~\eqref{eq:rate} are such that $\Vert \bF^\star \bG^\star \Vert = d$. Though the optimal solution to~\eqref{opt:qp} is not invariant to scaling, as far as the performance metric in~\eqref{eq:rate} is concerned,  there in no loss in optimality in scaling the solution given by BCD-SD, to fulfill the power constraint with equality. 
\end{remark}

\begin{table} 
\begin{algorithmic}
\Procedure{[$\bF$, $\bG$] $=$ BCD-SD }{$\tbGam_1$}  
\State Start with arbitrary $\bF_0$
\For {$k = 0, 1, 2, ... $}
    \State $\bG_{k+1} \leftarrow  (\bF_{k}^\dagger \bF_{k})^{-1} \bF_{k}^\dagger \tbGam_1 $ 
	\State $\bF_{k+1} \leftarrow  \Pi_{\calS}[\tbGam_1 \bG_{k+1}^\dagger (\bG_{k+1} \bG_{k+1}^\dagger)^{-1}]$ 
\EndFor
\EndProcedure
\end{algorithmic}
\vspace{.2cm}
\caption{Block Coordinate Descent for Subspace Decomposition (BCD-SD)} \label{alg:bcd_dcmp}
\end{table}

\subsubsection{One-dimensional case} \label{sec:sd1d}
Note that echoing (e.g., our proposed mechanism in Table~\ref{alg:distarn}) relies on the BS being able to send any vector $\bq_l$, to be echoed back by the MS. For the hybrid analog-digital architecture, this translates into the BS being able to (accurately) approximate $\bq_l$ by $\bff_l g_l$, where $\bff_l$ is a vector, $g_l$ is a scalar. As a result, subspace decomposition for the one-dimensional case is of great interest here. When $d=1$,~\eqref{opt:qp} reduces to the problem below, 
\begin{lemma} \label{lem:bfsol}
Consider the single dimension SD problem,
\begin{align} \label{opt:bfb}
\begin{cases}
               \underset{\bff, \ g}{\min} \ h_o(\bff, \ g) = \Vert \bff \Vert_2^2 \ g^2 -  2g\Re(\bff^\dagger \tbgam_1 )  \\
               \st  \ [\bff]_i = 1 /\sqrt{M} \ e^{j\phi_i} , \forall i  
            \end{cases}
\end{align}
where $g \in \mathbb{R}_+ $ and $[\tbgam_1]_i  = r_i e^{j \theta_i}$. Then the problem admits a globally optimum solution given by, $[\bff^\star]_i = 1 /\sqrt{M} \ e^{j\theta_i}, \forall i $ and $g^\star = \Vert \tbgam_1 \Vert_1 / \sqrt{M}$
\end{lemma}
\begin{IEEEproof} Refer to Appendix~\ref{app:3} \end{IEEEproof}
Similarly to~\eqref{eq:pc}, it can be verified that a power constraint is indeed implicitly verified. 
Moreover, the approximation error $\be \triangleq \tbgam_1 - \bff g$ is such that, 
\begin{align} \label{eq:dcmperr}
 [\be]_i  = \vert r_i - \Vert \tbgam_1 \Vert_1 /M \vert e^{j \theta_i}, \ \forall i \in \lbrace M \rbrace.
\end{align}
We note that when considering the effective beamformer, i.e., $\bff g$, the solution given by Lemma~\ref{lem:bfsol} is to some extent reminiscent of equal gain transmission in~\cite{Love_EGT_03, Zheng_EGT_07}, in terms of the optimal phases.  

We recall that a similar hybrid beamforming setup  was considered in~\cite{Zhang_VSP_05} where the authors optimize $ u , \bw ,\bff , g$, to maximize the  SNR as well as the spectral efficiency. Although  our formulation optimizes the same quantities, the optimization metric we consider, the subspace distance, is different. 

Note that the decomposition can be written in a  simple form. Given a vector $\tbgam_1$, its globally optimal decomposition (from the perspective of~\eqref{opt:qp}) is given as, 
$$ \tbgam_1 \approx
 g_1^\star \bff_1^\star  \triangleq (\Vert \tbgam_1 \Vert_1 / \sqrt{M}) \ \Pi_{\calS}[ \tbgam_1 ].$$ 
This can be generalized to obtain an alternate method to BCD-SD, by decomposing $\tbGam_1$, in a \emph{column-wise} fashion, 
\begin{align} \label{eq:cwdcmp}
 \tbGam_1 &= [ \tbgam_1, \cdots, \tbgam_d ] \approx [ g_1^\star \bff_1^\star, \cdots,  g_d^\star \bff_d^\star ] \nonumber \\
& \triangleq (1/\sqrt{M}) \ \left[ \Pi_{\calS}[ \tbgam_1 ], \cdots, \Pi_{\calS}[ \tbgam_d]  \right] \ \textrm{diag}( \Vert \tbgam_1 \Vert_1, \cdots, \Vert \tbgam_d \Vert_1  ) 
\end{align}

\subsubsection{Numerical Results}
\begin{figure}
	\center
	\includegraphics[width=9.5cm, height=7cm]{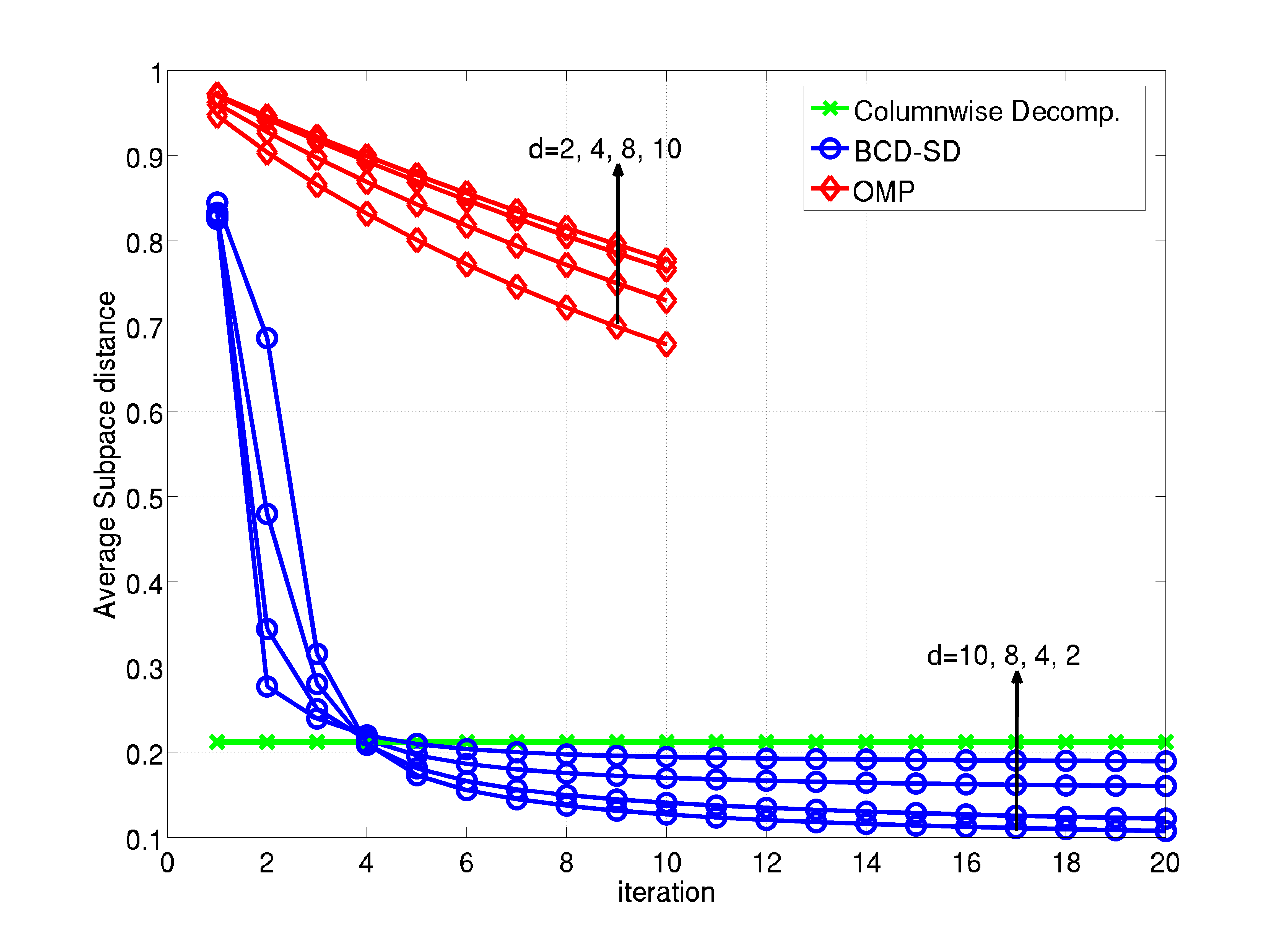}
	\vspace{.2cm}
	\caption{ Average subspace distance $\Vert \tbGam_1 - \bF \bG \Vert_F^2$, for our proposed method and OMP }
	\label{fig:dcmp}
\end{figure}

As mentioned earlier,~\eqref{opt:qp} was formulated and solved in~\cite{Ayach_Spatially_14}, using a variation on the well-known Orthogonal Matching Pursuit (OMP), by recovering $\bF$ in a greedy manner, then updating the estimate of $\bG$ in a least squares sense. We thus compare its average performance with our proposed method, for a case where $\tbGam_1 \in \bbC^{M \times d} $ is such that $M=64, r = 10$ (for several values of $d$). 
The curves are averaged over $500$ random realizations of $\bGam_1$ (the latter are random unitary matrices). Moreover, we follow the same setup for OMP as that of~\cite{Ayach_Spatially_14}, namely, that the dictionary is designed based on the array response vectors (of size $256$). 
  The reason for the large performance gap in Fig.~\ref{fig:dcmp} is that BCD-SD attempts to find a locally optimal solution to~\eqref{opt:qp} (though this cannot be shown due to the coupled variables). Moreover, OMP is halted after $r$ iterations, since it recovers the columns of $\bF$ one at a time, whereas our proposed method runs until reaching a stable point. 
With that in mind, although OMP  might perform better in terms of approximating the span of $\bGam_1$, it is challenging to measure and optimize such metrics in practice. Moreover, we recall that in its original formulation in~\cite{Ayach_Spatially_14} OMP is indeed formulated to solve the problem at hand (i.e.~\eqref{opt:qp}), and thus the comparison seems fair. 
 Interestingly, despite its extreme simplicity, the column-wise decomposition in~\eqref{eq:cwdcmp} offers a surprisingly good performance (as seen in Fig.~\ref{fig:dcmp}).

\subsection{Echoing in the Hybrid Analog-Digital Architecture}\label{sec:echohyb}
It is clear by now that the gist behind the schemes described in this work, is to obtain an estimate of $\lbrace  \bH^\dagger \bH \bq_l \rbrace_{l=1}^m $ at the BS, by exploiting channel reciprocity, using BS-initiated echoing described in ~\eqref{eq:echo}. However, in the case of the hybrid analog-digital architecture, there are several issues that prevent the application of the latter procedure. Firstly, the digital beamforming vector $\bq_l$ needs to be approximated by a cascade of analog and digital beamformer, using the decomposition in Sect.~\ref{sec:dcmp}, i.e., $\bq_l = \tbf_l \tilde{g}_l +\be_l $, where $\be_l$ is the approximation error given in~\eqref{eq:dcmperr}. Moreover, the  BS-initiated echoing relies on the MS being able to amplify-and-forward its received signal: this is clearly \emph{not possible} using the hybrid analog-digital architecture. In addition, neither the BS nor MS can digitally process the received signal at the antennas: only after the application the analog precoder/combiner (and possibly the digital precoder/combiner) can the baseband signal be digitally manipulated~\cite{Wang_Beamcodebook_09, Ayach_Spatially_14}. 

With this in mind, we \emph{emulate} the A-F step in BS-initiated echoing,~\eqref{eq:echo}, as follows.  $\bq_l$ is decomposed into $\tbf_l \tilde{g}_l$ at the BS and sent over the DL. The MS linearly processes the received signal in the downlink, with the analog combiner, i.e., $\bs_l = \bW_l^\dagger(\bH \tbf_l \tilde{g}_l )$, and same filter is used as the analog precoder, to process the transmit signal in the UL, i.e., $\bW_l \bs_l $. Finally, the received signal at the BS is processed with the analog precoder, $\bF_l$. The resulting estimate, $\bp_l$, at the BS is, 
\begin{align} \label{eq:echowrong}
\bp_l =\bF_l^\dagger \bH^\dagger \bW_l \bW_l^\dagger \bH (\bq_l - \be_l)
\end{align} 
Note that the above process is possible using the hybrid analog-digital architecture. 
Since noise is present in any uplink/downlink transmission, for clarity in what follows, we drop the noise-related terms from all equations. Needless to say, their effect is accounted for in the numerical results.
It is clear from~\eqref{eq:echowrong} that $\bp_l$ is no longer a ``good'' estimate of $\bH^\dagger \bH \bq_l$, for the reasons stated below.
\begin{itemize}
\item[1.] \emph{Analog-Processing Impairments (API):} Processing the signal at the MS with the analog combiner/precoder $\bW_l$ greatly distorts the singular values/vectors of the effective channel. Moreover, processing the received signal at the BS with the analog combiner $\bF_l \in \bbC^{M \times r} $ implies that $\bp_l$ is now a low-dimensional observation of the desired $M$-dimensional quantity $ \bH^\dagger \bH \bq_l$ (since $r < M$).
\item[2.] \emph{Decomposition-Induced Distortions (DID): } The error from decomposing $\bq_l$ at the BS, $\be_l$, further distorts the estimate (as seen in~\eqref{eq:echowrong}).
\end{itemize}
The above impairments are a byproduct of shifting the burden of digital precoding, to the analog domain. In what follows, these impairments will individually be investigated and addressed.

\subsubsection{Cancellation of Analog-Processing Impairments} \label{sec:api_full}
Our proposed method for mitigating analog-processing impairments (API) relies on the simple idea of taking multiple measurements at both the BS and MS, and linearly combining them, such that $\bW_l\bW_l^\dagger$ and $\bF_l \bF_l^\dagger$  approximate an identity matrix. 

In the DL, $\bq_l$ is approximated by $ \tbf_l \tilde{g}_l$, and  $\tbf_l \tilde{g}_l$ is sent over the DL channel\footnote{When sending $\tbf_l \tilde{g}_l$ over the DL, we can use $d$ RF chains, i.e., $$\bF_l \bG_l \ \pmb{1}_d = [\tbf_l, \cdots , \tbf_l] \ \textrm{diag}( \tilde{g}_l, \cdots , \tilde{g}_l) \ \pmb{1}_d = d \tbf_l \tilde{g}_l $$ thereby resulting in an array gain factor of $d$. Moreover, since we know from~\eqref{eq:pc} that $ \Vert \tbf_l \tilde{g}_l \Vert_2^2 \leq 1$, indeed this transmission scheme satisfies the power constraint. We also make use of this observation in the UL sounding.}, 
$K_r$ times (where $K_r = N/r$), each linearly processed with an analog combiner $\lbrace \bW_{l,k} \in \bbC^{N \times r} \rbrace_{k=1}^{K_r}$, to obtain the digital samples $\lbrace \bs_{l,k} \rbrace_{k=1}^{K_r}$ (this process is shown in Table~\eqref{alg:raid}). Moreover, the analog combiners  are taken from the columns of a Discrete Fourier Transform (DFT) matrix, i.e, 
\begin{align} \label{eq:dft1}
&[\bW_{l,1}, ..., \bW_{l,K_r}] = \bD_r, 
\end{align}
where $\bD_r \in \bbC^{N \times N}$ is a normalized $N \times N$ DFT matrix (i.e., where each column has unit norm and satisfies the unit-modulus constraint). The same analog combiners, $\lbrace \bW_{l,k} \rbrace_{k}$, are used to linearly combine $\lbrace \bs_{l,k} \rbrace_{k}$, to form $\tbs_l$ . We dub this procedure Repetition-Aided (RAID) Echoing, and the aforementioned DL phase, is shown in Table~\ref{alg:raid}. 
The resulting signal at the MS, $\tbs_l$, can be rewritten as,
\begin{align} \label{eq:tbs}
\tbs_l = \left( \sum_{k=1}^{K_r} \bW_{l,k} \bW_{l,k}^\dagger \right) \bH (d \tbf_l \tilde{g}_l) = d \bH  \tbf_l \tilde{g}_l ,
\end{align}
where equality follows from our earlier definition of $\lbrace \bW_{l,k}  \rbrace_k$ in~\eqref{eq:dft1}. Note that \emph{the effect of processing the received signal with the analog combiner has been completely suppressed}. Now, $\tbs_l$ is normalized, and echoed back in the UL direction. 

A quite similar process is used in the UL:  $\tbs_l$ is first decomposed into $\tbw_l \tilde{u}_l$, $d$ RF chains are used to send it over the UL, $K_t$ times (where $K_t = M/r$), and each observation is linearly processed with an analog combiner $\lbrace \bF_{l,m} \in \bbC^{M \times r} \rbrace_{m=1}^{K_t}$. The resulting digital samples $\lbrace \bz_{l,m} \rbrace_{m=1}^{K_t}$ are again linearly combined with the same $\lbrace \bF_{l,m}  \rbrace_m$, to obtain the desired estimate $\bp_l$. Similar to the DL case, the analog combiners  are taken from the columns of a Discrete Fourier Transform (DFT) matrix, i.e, $[\bF_{l,1}, ..., \bF_{l,K_t}] = \bD_t $. The process for the UL is also shown in Table~\ref{alg:raid}. We combine its steps to rewrite $\bp_l$ as, 
\begin{align} \label{eq:bp}
\bp_l &= \left( \sum_{m=1}^{K_t} \bF_{l,m} \bF_{l,m}^\dagger \right) \bH^\dagger (d \tbw_l \tilde{u}_l ) = d \bH^\dagger  \tbw_l \tilde{u}_l 
\end{align}
\begin{figure}
	\center
	\includegraphics[trim=0cm 0cm 0cm 4cm, clip=true, scale=.6 ]{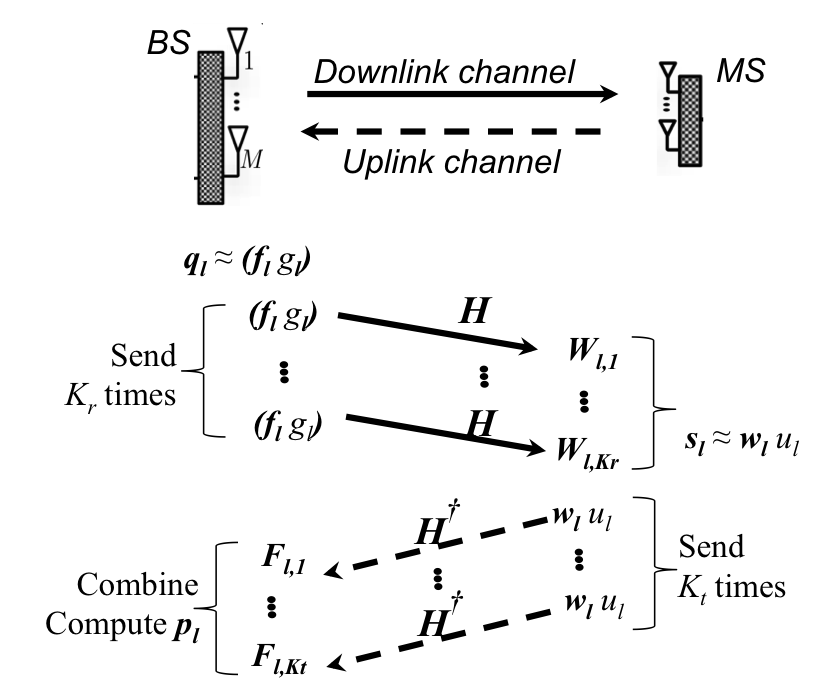}
	\vspace{.2cm}
	\caption{Repetition-aided (RAID) echoing for the hybrid analog-digital architecture}   
	\label{fig:echo}
	\vspace{-2em}
\end{figure}

At the output of the RAID procedure, the BS has the following $\bp_l$,
\begin{align}
\bp_l &= d \bH^\dagger \tbw_l \tilde{u}_l = d \bH^\dagger (\tbs_l - \be^{(r)}_l ) =  d \bH^\dagger (d \bH \tbf_l \tilde{g}_l - \be^{(r)}_l )   \nonumber   \\
&= d^2 \bH^\dagger \bH \bq_l - d^2 \bH^\dagger \bH \be^{(t)}_l - d \bH^\dagger \be^{(r)}_l   \label{eq:pl}
\end{align}
Note that $\be_l^{(t)}  =   \bq_l - \tbf_l \tilde{g}_l$ (resp. $\be_l^{(r)}  =   \tbs_l - \tbw_l \tilde{u}_l$) is the error emanating from approximating $\bq_l$ (resp. $\tbs_l$) at the BS (resp. MS), that we dub \emph{BS-side} (resp. MS-side) \emph{decomposition-induced distortion (DID)}. It is quite insightful to compare $\bp_l$ in  the latter equation with~\eqref{eq:echowrong}. We can clearly see that impairments originating from processing the received signals with both $\bW_l$ and $\bF_l$, have completely been suppressed. In~\eqref{eq:pl}, $\bp_l$ indeed is the desired estimate, i.e., $\bH^\dagger \bH \bq_l$, corrupted by distortions emanating from the BS-side decomposition, $\be^{(t)}_l$, and the MS side decomposition, $\be^{(r)}_l$ (both investigated later in the next subsection). Both UL and DL phases of he process are illustrated in Fig.~\ref{fig:echo}, and detailed in Table~\ref{alg:raid}.

\begin{remark} \rm
Note that employing this process reduces the hybrid analog-digital architecture into a conventional MIMO channel: any transmitted vector in the DL, $(\tbf_l \tilde{g}_l) $, can be received in a ``MIMO-like'' fashion, as seen from~\eqref{eq:tbs}, at a cost of $K_r$ channel uses (the same holds for the UL, as seen from~\eqref{eq:bp} ).  
\end{remark}
It can be seen from the above, that in the DL (resp. UL), $d$ RF chains are active at the BS (resp. MS), while all $r$ RF chains are used at the MS (resp. BS), to minimize the overhead. With this in mind, it can be seen that the associated overhead with each echoing,  $\Omega = (M+N)/r$ (channel uses), will decrease as more RF chains are used. 

\begin{table} 
\begin{algorithmic}
\State // \emph{DL phase}
\State \hspace{.5cm} $\bq_l = \tbf_l \tilde{g}_l + \be^{(t)}_l$
\State \hspace{.5cm} $\bs_{l,k} = \bW_{l,k}^\dagger \bH (d \tbf_l \tilde{g}_l), \ \forall k \in \lbrace K_r \rbrace$
\State \hspace{.5cm} $\tbs_l = \sum_{k=1}^{K_r} \bW_{l,k} \bs_{l,k}$
\State // \emph{UL phase}
\State \hspace{.5cm} $\tbs_l = \tbw_l \tilde{u}_l + \be^{(r)}_l $
\State \hspace{.5cm} $\bz_{l,m} = \bF_{l,m}^\dagger \bH^\dagger (d \tbw_l \tilde{u}_l ), \ \forall m \in \lbrace K_t \rbrace  $
\State \hspace{.5cm} $\bp_l = \sum_{m=1}^{K_t} \bF_{l,m} \bz_{l,m}   $
\end{algorithmic}
\vspace{.2cm}
\caption{Repetition-Aided (RAID) echoing} \label{alg:raid}
\end{table}

\subsubsection{Imperfect Compensation of Analog-Processing Impairments} \label{sec:api_imp}
Though the above method perfectly removes all artifacts of analog processing, the overhead is proportional to $(M + N)/r$. A natural question is whether a similar result can still be achieved when $\bD_r $ and $\bD_t$ are truncated matrices i.e. when $K_r < N/r$ and $K_t < M/r$. Perfect cancellation of API relies on a careful choice of the analog precoder/combiner for each measurement, by picking $\lbrace \bW_{l,k} \rbrace_{k=1}^{K_r}$ and $\lbrace \bF_{l,m} \rbrace_{m=1}^{K_t}$ to span all the columns of (square) DFT matrices. We investigate the effect of picking $\bD_r $ and $\bD_t$ as truncated matrices, i.e. when $K_r < N/r$ and $K_t < M/r$. Focusing our discussion on just analog precoders for brevity, we seek to find a (tall) matrix $\tbD_t \in \bbC^{M \times (\eta M)} $, $\eta < 1$, such that,
\begin{align} \label{opt:Dt}
\begin{cases}
               \underset{ \tbD_t }{\min} \Vert \frac{1}{ M} \bI_M - \tbD_t \tbD_t^\dagger  \Vert_F^2   \\
               \st \ \ \tbD_t \in \calS_{M, \ \eta M} \ . 
\end{cases}
\end{align}
Due to the apparent difficulty of the problem, one can resort to \emph{stochastic optimization} tools, e.g. simulated annealing: this approach is ideal for the design of  $\tbD_t$ (and $\tbD_r$ as well), since it is completely independent of all parameters (except $M, N$ and $\eta$), and can thus be computed off-line and stored for later use. Then, the resulting overhead would be reduced to $\Omega = \eta \frac{M+N}{r} $. Further investigations along this line are outside the scope of this work, but we opted to include them briefly, for completeness.

\subsection{Proposed Algorithms}
Combining the results of the previous subsections, we can now formulate our algorithm for Subspace Estimation and Decomposition (SED) for the hybrid analog-digital architecture (shown in Algorithm~\ref{alg:bsehyb}): estimates of the right / left singular subspaces, $\tbGam_1$ and $\tbPhi_1$, can be obtained by using the SE-ARN procedure (Sect.~\ref{sec:bsemimo}), keeping in mind that the \emph{echoing phase (Steps 1.a and 1.b) is now replaced by the RAID echoing procedure} (Table~\ref{alg:raid}. Then, the multi-dimensional subspace decomposition procedure, BCD-SD in Sect.~\ref{sec:dcmp}, is then used to approximate each of the estimated singular spaces, by a cascade of analog and digital precoder/combiner. 
We highlight a desirable feature for the SED algorithm: the subspace estimation mechanism is totally decoupled from the subspace decomposition part, and thus any of the latter parts can be substituted, if desired.

\begin{algorithm} 
\caption{Subspace Estimation and Decomposition (SED) for Hybrid Analog-Digital Architecture} \label{alg:bsehyb}
\begin{algorithmic}
\State // \emph{Estimate $\tbGam_1$ and $\tbPhi_1$ }
\State $\tbGam_1, \tilde{\bSig}_1 =$ SE-ARN ($\bH$, $d$) 
\State $\tbPhi_1 =$ SE-ARN ($\bH^\dagger$, $d$) 
\State // \emph{Decompose $\tbGam_1$ and $\tbPhi_1$ }
\State [$\bF$, $\bG$ ] = BCD-SD ($\tbGam_1$, $\rho$) 
\State [$\bW$, $\bU$ ] = BCD-SD ($\tbPhi_1$, $\rho$) 
\State Perform waterfilling on $\tilde{\bSig}_1$
\end{algorithmic}
\end{algorithm} 


Note that previously proposed algorithms within this context such as the PM and TQR in~\cite{Dahl_blind_04}, are no longer applicable here: indeed both rely on the MS being able to amplify-and-forward its received signal at the antennas - clearly this modus operandi cannot be supported by the hybrid analog-digital architecture. Interestingly, it is possible to apply elements from the RAID echoing structure that we developed, effectively modifying the original echoing structure of the latter schemes, and adapting them to the hybrid analog-digital architecture (as shown in Algorithm~\ref{alg:m2qr}). 

\begin{algorithm} 
\caption{Modified Two-way QR (MTQR) for Hybrid Analog-Digital Architecture} \label{alg:m2qr}
\begin{algorithmic}
\For{$l=1,2,..., I$}
\State // \emph{Decompose each column of $\bX$}
\State \hspace{.2cm} $[\bX]_n \approx \tbf_{n} \tilde{g}_{n}, \ \forall n \in \lbrace d \rbrace $ (using Lemma~\ref{lem:bfsol})
\State \hspace{.2cm} $\tbX = [\ \tbf_{1} \tilde{g}_{1} \ , \cdots , \ \tbf_{d} \tilde{g}_{d} \ ] $
\State // \emph{Send $\tbX$ in DL, one column at a time} 	
\State \hspace{.2cm} $\bT_{k} = \bW_{k}^\dagger \bH \tbX, \ \forall k \in \lbrace K_r \rbrace $ 
\State \hspace{.2cm} $\bY = \sum_{k=1}^{K_r} \bW_{k} \bT_{k} $ ; $\bY = $ qr($\bY$)
\State // \emph{Decompose of $\bY$}
\State \hspace{.2cm} $[\bY]_n \approx \tbw_{n} \tilde{u}_{n}, \ \forall n \in \lbrace d \rbrace $ (using Lemma~\ref{lem:bfsol})
\State \hspace{.2cm} $ \ \tbY = [\ \tbw_{1} \tilde{u}_{1} \ , \cdots , \ \tbw_{d} \tilde{u}_{d} \ ] $ 
\State // \emph{Send $\tbY$ in UL, one column at a time} 
\State \hspace{.2cm} $\bS_{k} = \bF_{k}^\dagger \bH^\dagger \tbY, \ \forall k \in \lbrace K_t \rbrace $ 
\State \hspace{.2cm} $\bZ = \sum_{k=1}^{K_t} \bF_{k} \bS_{k} $ ; $\bX = $ qr($\bZ$)
\EndFor 
\end{algorithmic}
\end{algorithm} 

Operationally, the proposed MTQR  algorithm is the same as the Two-way QR (TQR) in~\cite{Dahl_blind_04}, whereby  $\bGam_1$ and $\bPhi_1$ are obtained iteratively: as $I \rightarrow \infty$, $\bX \rightarrow \bGam_1 $ (at BS) and $\bY \rightarrow \bPhi_1 $ (at MS). At each iteration of the algorithm, the BS sends $\bX$ in the downlink, and the QR algorithm is applied to the received signal. Then, the resulting signal is sent by the MS in the uplink, and the QR algorithm is applied at the BS to form $\bY$. While TQR assumes fully digital MIMO transmission, our contribution is to apply the RAID scheme, to make the transmission compatible with the hybrid analog-digital systems.

\subsection{Bounds on Eigenvalue Perturbation}
It can be clearly seen that \emph{the iterative nature of Algorithm~\ref{alg:m2qr} makes the application of Lemma~\ref{lem:pert}, to quantify the impact of decomposition and approximation errors, not possible}. On the other hand, for Algorithm~\ref{alg:bsehyb}, the fact that each $\bH^\dagger \bH \bq_l $ is only corrupted by two sources of DID, $\be_l^{(r)}$ and $\be_l^{(r)}$, makes the latter possible. With that in mind, \emph{we specialize the result of Sect.~\ref{sec:mimobse} and Lemma~\ref{lem:pert} (developed for generic MIMO systems) to the case of Algorithm~\ref{alg:bsehyb} in the hybrid analog-digital architecture}. We thus relate the eigenvalues of $\bT_m$ at the output of SE-ARN, to the dominant eigenvalues of $\bC_m$, and consequently of $\bA$ (Sect.\ref{sec:mimobse}). 

\begin{corollary} \label{cor:pert}
Every eigenvalue $\tilde{\lambda}$ of $\bT_m$ satisfies 
\begin{align*}
\vert \tilde{\lambda} - \lambda \vert \leq  m  \Vert \bH \Vert_F^2  \ (3 +  \frac{1}{d \Vert \bH \Vert_F}  )
\end{align*}
where $\lambda$ in an eigenvalue of $\bC_m$.
\end{corollary}
\begin{IEEEproof}   Refer to Appendix~\ref{app:5}
\end{IEEEproof}
Moreover, recall that as $m \rightarrow M$, $\lambda$ is an eigenvalue of $\bA$ (Lemma~\ref{lem:arnlem} - P3). Thus, this result directly relates the eigenvalues of $\bT_m$, to that of $\bA$: though this holds asymptotically in $m$, our simulations will show that good approximations can still be obtained, even for $m \ll M$. Note that we have ignored the effect of DID compensation, within the RAID echoing process, for convenience. As a result, the above bound is a  ``pessimistic'' performance measure. 

\subsection{Practical Implementation Aspects}
We evaluate  the \emph{communication overhead} of both schemes, in number of channel uses, keeping in mind that the actual overhead will be dominated by the latter.  Algorithm~\ref{alg:bsehyb} requires $K_t+K_r$ channel uses per iteration, to estimate $\tbGam_1$, and $K_t+K_r$ to estimate $\tbPhi_1$, for a total of 
\begin{align} \label{eq:ovh_arn}
\Omega_{SED} = 2m \ \frac{M+N}{ r } ,    
\end{align}
$m$ being the number of iterations for the Arnoldi process. Letting $I$ denote the number of iterations for MTQR, the number of channel uses required for Algorithm~\ref{alg:m2qr} is, 
\begin{align}
\Omega_{MTQR} = d I   \ \frac{M+N}{r }    
\end{align}

It should be emphasized here that our main focus in this work is to investigate the principle of subspace estimation employing numerical techniques, and through simulations describe the performance gain that can be expected by taking on such an approach. Hence, our major concern is not to investigate a stable and low-complexity technique that can be readily implemented in practice. We will, however, provide suggestions on what can be done to enhance the stability of the devised schemes, while admitting that many of the problems connected with practical implementation of the proposed method are subject to further study.
Generally, it is known that the Arnoldi (and Lanczos) algorithm may suffer from numerical stability issues. Though analytically speaking, the basis $\bQ_m$ is easily shown to be orthonormal, in practice, however, errors resulting from floating-point operations lead to a loss in orthogonality (the extent to which it happens is dependent on the application)~\cite[Sec. 7.3]{Saad_Numerical_11}. Moreover, for our algorithm, noise inherent to the echoing process will further amplify this effect. 
One of the widely adopted fixes for this matter is the Implicitly Restarted  Arnoldi algorithm~\cite[Sec. 7.3]{Saad_Numerical_11}. We did experiment with such an algorithm, and though it does enhance the numerical  stability of the algorithm, the resulting overhead is increased by a large factor. This issue is critical for the SED algorithm (that employs the RAID echoing), since it renders real-world implementation quite impractical. Moreover, there are many problems connected with practical implementations of the Restarted Arnoldi method, that are subject to further study. Other methods that might enhance the stability the Arnoldi Iteration, such as deflation techniques, have been reported in~\cite{Sorensen_IRA_96}. 

\subsection{Discussion} \label{sec:disc}
We have presented an approach to maximizing the metric $R$ defined in~\eqref{eq:rate}. As mentioned earlier, the value of the objective function is in general \emph{not an achievable rate} for our system. However, optimizing similar expressions related to achievable rates has been proved to give good
results in previous work on transmission with partial CSI~\cite{Baum_MIMOsound_11}. Since any rate achievable with partial CSI, cannot be larger than the corresponding rate achievable with perfect CSI, this criterion always provides an upper bound on the achievable rates in our system. Hence, in our approach, if the proposed algorithms result in values for $R$ that are closing in on the perfect CSI upper bound, then the scheme is performing optimally (in the sense of achievable rates).

With the above in mind, we use the following, as our performance metric in the simulations,
\begin{align}
\tilde{R} = \log_2 \left| \bI_d + \frac{1}{ \sigma_{(r)}^2} \bU^\dagger \bW^\dagger \bH \bF \bG \bG^\dagger \bF^\dagger \bH^\dagger \bW \bU  (\bU^\dagger \bW^\dagger \bW \bU)^{-1}\right| \ .
\end{align}
In that sense, $\tilde{R} $ is the `user rate' that is based on the actual channel $\bH$, and the precoders / combiners that are in turn designed based on the estimated channel.

\section{Numerical Results}
\subsection{Simulation Setup }
In this section, we numerically evaluate the performance of our algorithms, in the context of a single-user MIMO link. We adopt the prevalent physical representation of sparse mmWave channels adopted in the literature, e.g.,~\cite{Alkhateeb_channel_2014, Ayach_Spatially_14}, where  only $L$ scatterers are assumed to contribute to the received signal - an inherent property of the poor scattering nature in mmWave channels, 
\begin{align} \label{eq:chann}
\bH = \sqrt{\frac{M N}{L	}} \sum_{i=1}^L \beta_i  \ \ba_r(\chi_i^{(r)}) \ba_t^\dagger(\chi_i^{(t)} )
\end{align}
where $\chi_i^{(r)}  $ and $\chi_i^{(t)}$ are angles of arrival at the MS, and angles of departure at the BS (AoA/AoD) of the $i^{th}$ path,  respectively (both assumed to be uniform over $[ -\pi/2 , \ \pi/2]$), $\beta_i$ is the complex gain of the $i^{th}$ path such that $\beta_i \sim \mathcal{CN}(0, 1), \  \forall i $. Finally, $\ba_r(\chi_i^{(r)})$ and $\ba_t(\chi_i^{(t)})$ are the array response vectors at both the MS and BS, respectively. For simplicity, we will use uniform linear arrays (ULAs),  where we assume that the inter-element spacing is equal to half of the wavelength.  In what follows, we also assume that $M/r = 8$ and $N/r = 4$, i.e., as $M,N$ increase, so does the number of RF chains. 

\subsubsection{Benchmarks/Upper bounds}
We use the Adaptive Channel Estimation (ACE) method (Algorithm 2 in~\cite{Alkhateeb_channel_2014}) as a benchmark, to estimate the mmWave channel. It is based on sounding of \emph{hierarchical codebooks} at the BS, feedback of the best codebook indexes by the MS, and finding the analog/digital precoders and combiners using OMP~\cite{Ayach_Spatially_14}.  Moreover, the authors characterized the resulting communication overhead $\Omega_{ACE}$, as a function of the codebook resolution. We used the corresponding MATLAB implementation that was provided by the authors. We adjust the number of iterations for both our proposed schemes and the codebook resolution of benchmark scheme, such that $\Omega_{SED} = \Omega_{MTQR} \triangleq \Omega_o  \approx \Omega_{ACE} $. Note that we do not assume any quantization for phases of the RF filters. 
We also compare the performance of the algorithms against the ``optimal performance'', $R^\star$ in~\eqref{eq:optrate}, where full CSIT/CSIR is assumed, fully digital precoding is employed, and the optimal precoders are used.  All curves are averaged over $500$ channel realizations. 

\begin{remark} \rm
Note that if one want to use ``classical'' pilot-based channel estimation to estimate the DL channel, i.e., a pilot sequence of minimum length $M$, then the same repetition-based framework that was used in RAID echoing, has to be used to cancel the effect of $\bW$ from the effective channel estimate: it can be easily seen that the resulting total (both DL and UL) number of pilots slots would be $2MN/r^2$, thereby making the latter method infeasible. 
\end{remark}

\subsection{Performance Evaluation}
  
\begin{figure}
    \center
	\includegraphics[width=9.5cm, height=7cm]{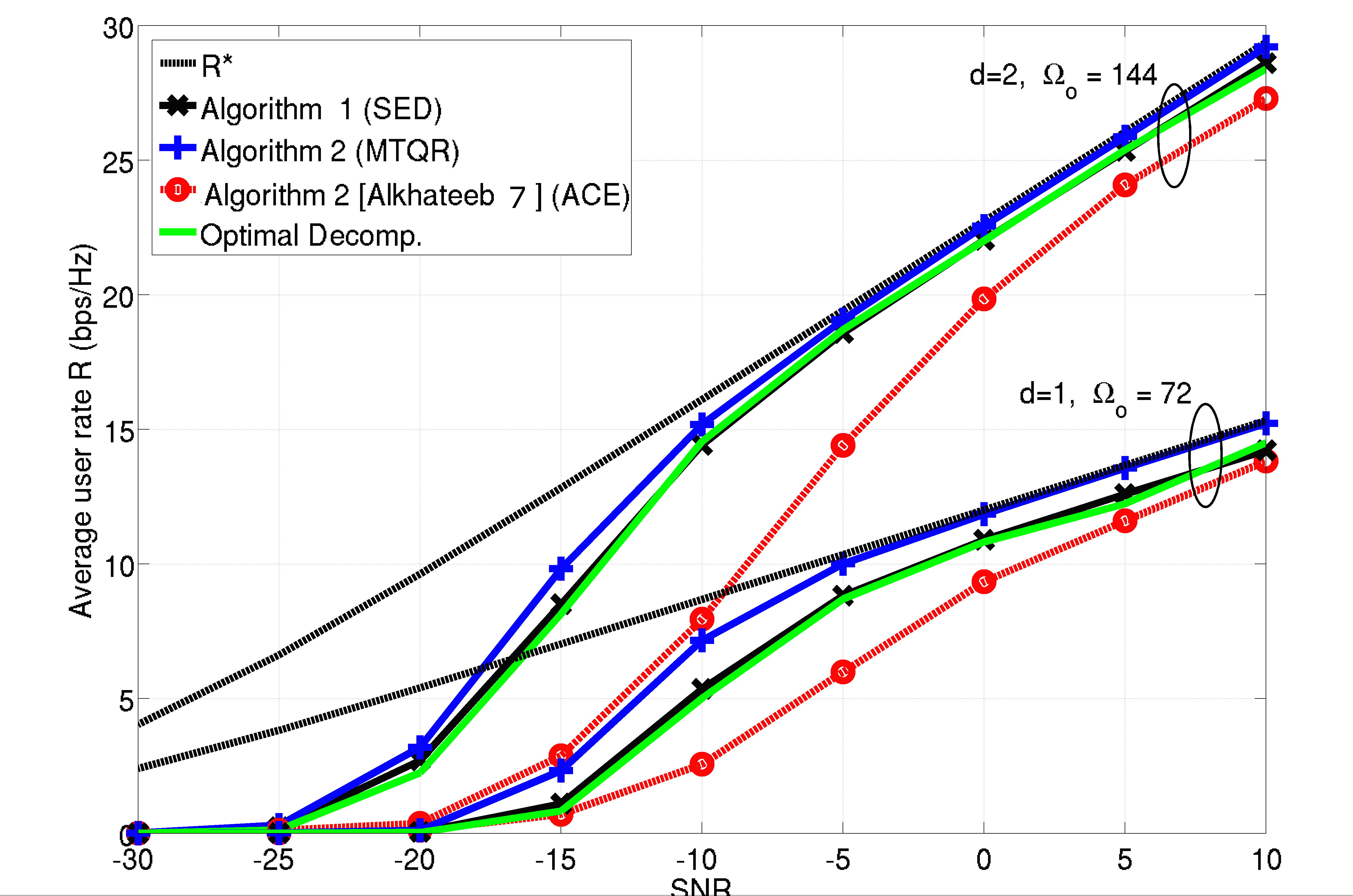}
	\vspace{.1cm}
	\caption{  Average sum-rate of proposed schemes ($M=128, N=64, d=2, L=3, m =6$)}  
    \label{fig:sr1}
	
    \center
	\includegraphics[width=9.5cm, height=6.5cm]{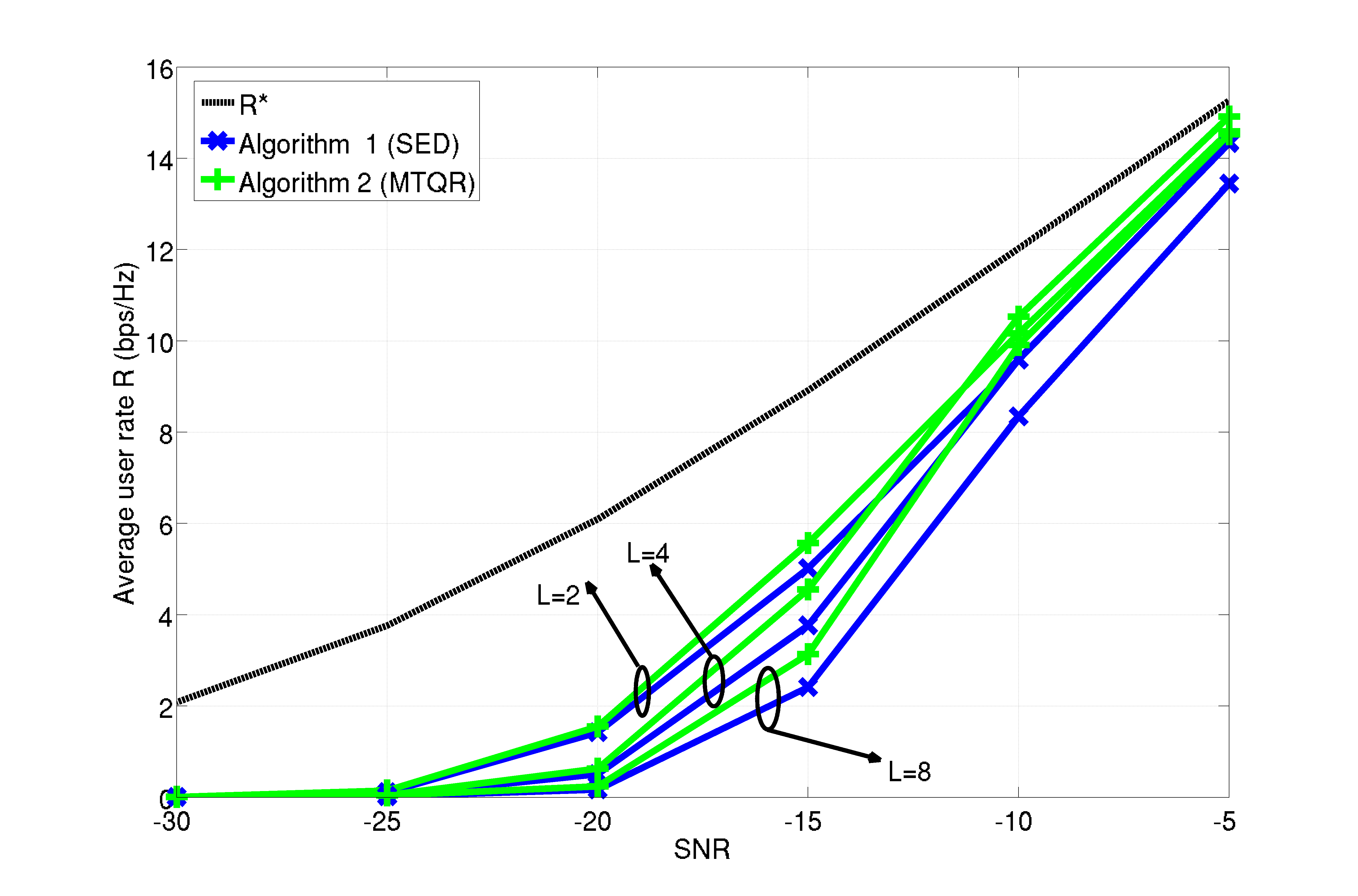}
	\caption{  Effect of number of paths L, on the average user rate ($M=64, N=32, d=2, m = 6$) } 
	\label{fig:srvsL}

	    \center
	\includegraphics[width=10cm, height=4.5cm]{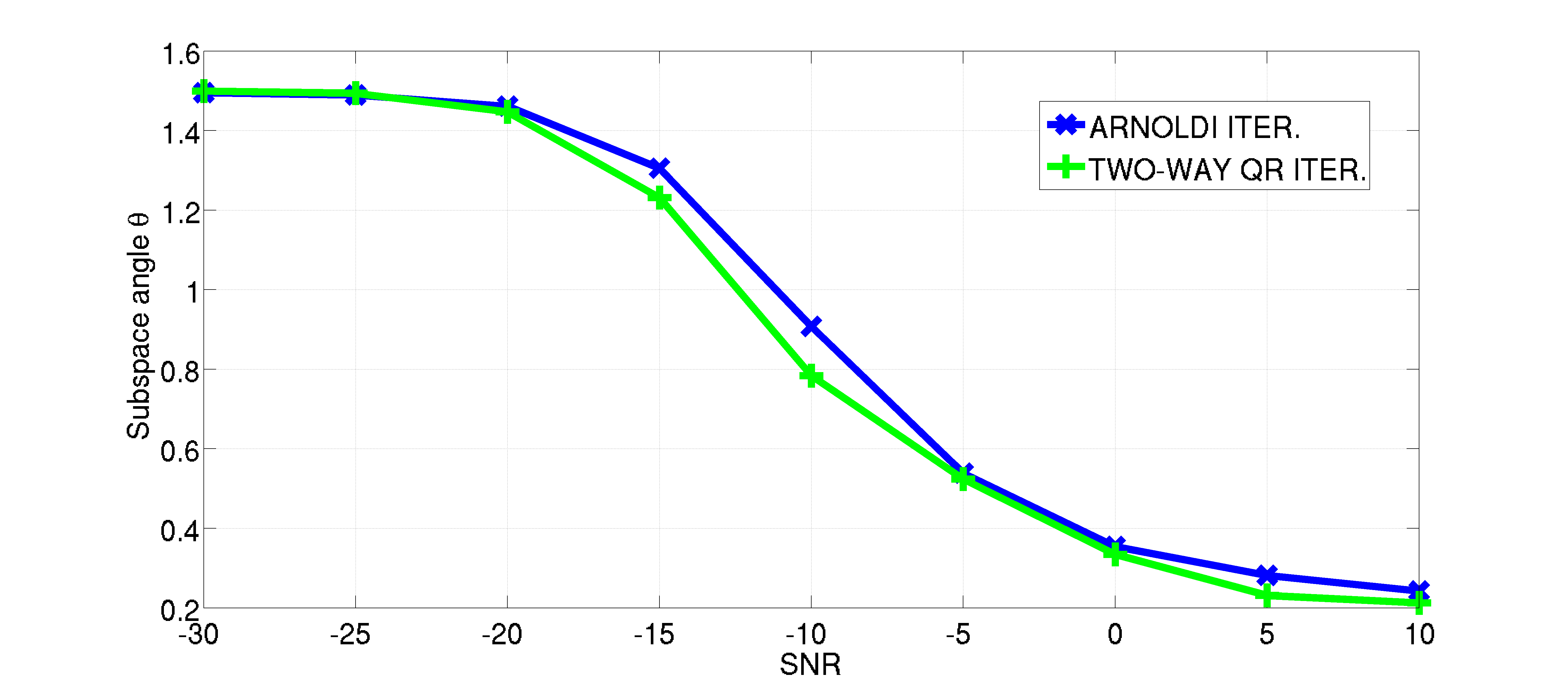}
	\vspace{.1cm}
	\caption{  Average subspace angle ($M=64, N=32, d=3, L=4, m =6$)} 
    \label{fig:avg_dist}
\end{figure}

We start by investigating the performance of our schemes against the above benchmarks, for the case where $M=128, N=64, L=3$, and  $m =3d$, for two cases: $d=1$ and $d=2$ where the resulting overhead is $\Omega_o = 72$ and $\Omega_o = 144$ channel uses, respectively. It can be seen from Fig.~\ref{fig:sr1} that both \emph{proposed schemes exhibit relatively similar performances}, that are in turn very \emph{close to the optimal performance bound $R^\star$} (especially above $-10$ dB). This indeed suggests that the multiplexing gain achieved by conventional MIMO systems can still be maintained in the hybrid analog-digital architecture, albeit at a much lower cost: the number of \emph{required RF chains can be drastically decreased}, resulting in savings in terms of cost and power consumption. Moreover, we observe a sharp and  significant performance gap between both our schemes and the benchmark from~\cite{Alkhateeb_channel_2014}, over all SNR ranges (the gap being more significant in the low-SNR regime). 
We also evaluate the so-called optimal decomposition schemes~\cite{Zhang_VSP_05, Nuria_hybrid_15} that can exactly decompose $\bGam_1$ into $\bF \bG$ (discussed in Sec. IV). Thus, the curves labeled 'Optimal Decomp.' refer to the case where the optimal decomposition is used in conjunction with SED. Fig~\ref{fig:sr1} clearly reveals that the ability to optimally decompose the estimated subspaces does not bring about additional gains. We note that the tiny mismatch between 'Optimal Decomp.' and Algorithm 1 is due to simulation resolution.

We attempt to shed light on the stability of the proposed algorithms, as the number of paths in the mmWave channel, $L$, increases (where we set $M=64, N=32, d=2, m=6$). For clarity we restrict the result to the low SNR regime. Though a degradation in the performance of both algorithms is expected, as $L$ increases, Fig.~\ref{fig:srvsL} clearly indicates that the latter degradation is not quite significant. Though not visible here, our simulations show that this degradation is not present in the medium-to-high SNR region. As expected, this technique is best used for channels with a few paths, e.g., mmWave channels.

We investigate the performance of both SED and MTQR in terms of average subspace angle,  $\theta = \mathbb{E}[\alpha(\bGam_1, \tbGam_1 ) ]$ where $\alpha(\bGam_1, \tbGam_1 )$ (radians) is defined as the subspace angle between $\bGam_1$ and $\tbGam_1$ (implemented by computing the principal angles of the latter subspaces). As shown in Fig.~\ref{fig:avg_dist}, both schemes exhibit a similar behavior of better estimation accuracy, as the SNR increases.

\begin{remark}  \label{rem:sed_vs_mtqr} \rm 
Though the performance of Algorithm~\ref{alg:m2qr} seems to be better, Fig.~\ref{fig:sr1}-\ref{fig:avg_dist} both suggest that this gap is quite narrow. Moreover, both algorithms seem to exhibit very similar behavior. With that in mind, and for the sake of clarify of our results, we opt to focus on Algorithm~\ref{alg:bsehyb}, the main object of investigation in this work.  
\end{remark}

We next investigate its scalability: we scale up $M$ and $N$ (assuming $N= M/2$, for simplicity), while keeping everything else fixed, i.e., $d=2 , m=6$, and consequently $\Omega_o = 144$. In doing that, we noticed that the complexity of the benchmark scheme~\cite{Alkhateeb_channel_2014} was \emph{prohibitively high}, thus preventing us from investigating its scalability: we were unable to get any results for systems larger than $128 \times 64$ . On the other hand, both our algorithms exhibit no such problems since all the computations that they involve are matrix-vectors/matrix-matrix operations. Consequently, the \emph{complexity gap between Algorithm~\ref{alg:bsehyb} and the benchmark increases drastically, as $M,N$} grow. 

\begin{figure}[!]
	\center
	\includegraphics[width=9.5cm, height=6.5cm]{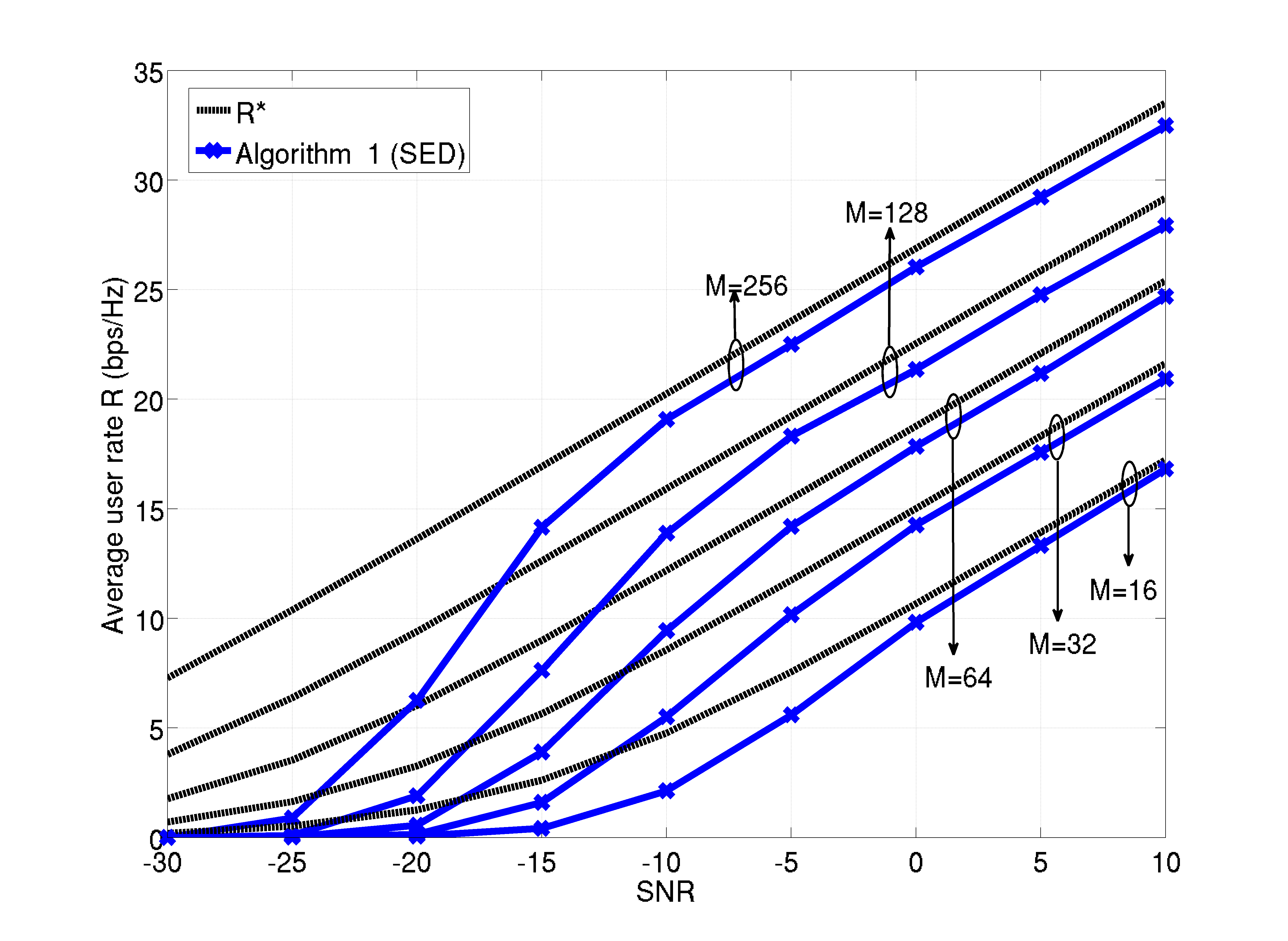}
	\vspace{.1cm}
	\caption{Average user-rate for different $M,N$ ($N=M/2, d=2, L=4, m =6, \Omega_o = 144$) }
	\label{fig:sr3}		
	\center
	\includegraphics[width=9.5cm, height=7.5cm]{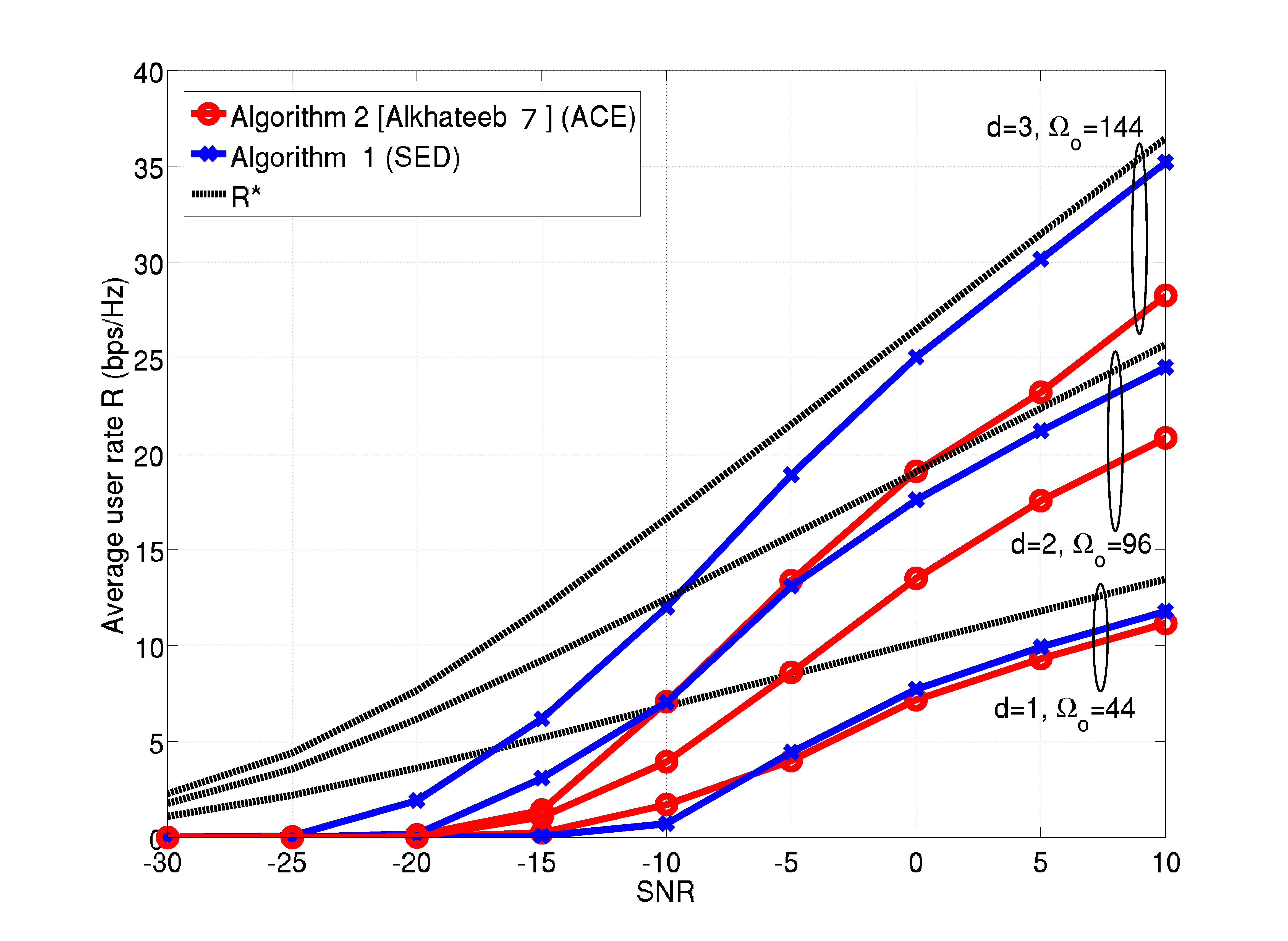}
	\vspace{.1cm}
	\caption{Average user-rate of proposed schemes over SCM channels ($M=64, N=32, m = 2d$) }
	\label{fig:sr2}
\end{figure}

Fig~\ref{fig:sr3} clearly shows that Algorithm~\ref{alg:bsehyb} is able to harness the significant array gain inherent to large antenna systems (by closely following the optimal performance bound, $R^\star$, with a small constant gap), while keeping the overhead remarkably small. Though the performance might not be good enough to offset the overhead, for the $16 \times 8$ case, it surely does for the $256 \times 128$. Moreover, note that the gap between the optimal performance and Algorithm~\ref{alg:bsehyb} is quite small (across the entire SNR range) for small systems dimensions, and quite small even for large values of $M$ (at high SNR).  The key to this result is to have $M/r$ and $N/r$ fixed, as $M,N$ increase.

We also evaluate the performance of Algorithm~\ref{alg:bsehyb} in a more realistic manner, by adopting the Spatial Channel Model (SCM) detailed in~\cite{3GPP_tr25996, SCM_Matlab}, and modifying its parameters to emulate mmWave channels: the number of paths is set to $4$, the carrier frequency to $60$ GHz, the BS/MS array is modified to implement ULAs, and an 'urban micro' scenario is selected, where a small $\Omega_o$ is desired. Fig.~\ref{fig:sr2} shows the average performance of such a system, with $M=64, N=32, m = 2d$, for several values of $d$ (each resulting in different values for $\Omega_o$). Though both our algorithm and the benchmark exhibit similar performances for $d=1$, this gap increases with $d$, e.g. for $d=3$ this performance gap is quite significant. Moreover, we can clearly see that Algorithm~\ref{alg:bsehyb} yields a relatively high throughput in this realistic simulation setting (especially for $d=3$), while still keeping the overhead at a relatively low level. 

Evidently, increasing $m$ (the number of iterations for the Arnoldi) has the effect enhancing the estimation accuracy (and increasing the communication overhead as well~\eqref{eq:ovh_arn}). The marginal improvement brought about by increasing $m$, is decreasing, and thus our simulations indicated that setting $ 2d \leq  m \leq 3d $ provides a good trade-off.

\subsection{Discussions}

A few remarks are in order at this stage, regarding similarities and differences between our two proposed algorithms. As discussed in Remark~\ref{rem:sed_vs_mtqr}, when the communication overhead is normalized, both SED and MTQR exhibit a similar behavior and performance profile, across the entire SNR range (with a relatively small performance gap): indeed they can be used interchangeably with no change at all in the operational requirements. However, as this work shows, we have an accurate analytical description of the behavior of SED: the Arnoldi algorithm was adapted to the  subspace estimation part (with some analytical performance guarantees), and BCD-SD to mathematically describe the decomposition algorithm. In contrast, MTQR is a (heuristic) variation on the original TQR, whose behavior we have not modeled analytically. 

One of the conclusions suggested by all the above results, is the fact that the low-SNR performance of the proposed schemes is rather poor. However, interestingly,  Figs.~\ref{fig:sr1}-\ref{fig:sr2} unambiguously point out that this is the case for the benchmark scheme as well (ACE in~\cite{Ayach_Spatially_14}): one might be tempted to conjecture at this point that this low-SNR behavior is an inherent aspect of mmWave channel estimation. 
Initial investigations reveal that, if more RF chains (more than $r$) can be employed during the RAID echoing phase, the low-SNR performance can be greatly boosted. 

\section{Conclusion}
We proposed an algorithm for blindly estimating the left and right singular subspace of a mmWave MIMO channel, by exploiting channel reciprocity that is inherent to TDD systems. Though the algorithm is a perfect match for conventional (large) MIMO systems, we extended it to operate under the constraints dictated by the hybrid analog-digital architecture, and showed via simulations that it is a good fit for large MIMO channels, with low rank, e.g., mmWave channels. Finally, our simulations showed that a similar performance to the ideal case (fully digital perfect CSI) can be achieved, with a only a few RF chains, thereby resulting in significant saving in energy and cost, over conventional MIMO systems. 

\section{Acknowledgments}
We are deeply indebted to the reviewers, whose invaluable comments greatly improved the manuscript.    

\appendix
\section{}

\subsection{Proof of Lemma~\ref{lem:arnlem} } \label{app:1}
\noindent $(P1):$ 
Combining steps (2.b) and (3.a) in the SE-ARN procedure, we write, 
\begin{align*}
\bA \bq_l + \tbw_l = \sum_{i=1}^{l+1}  [\tbT_m]_{i,l} \ \bq_i + \sum_{i=1}^l  [\bE_m]_{i,l}  \ \bq_i \ , \ \forall l \in \lbrace m \rbrace,
\end{align*}
We can rewrite the latter equation in matrix form, using the definitions of  $\tbT_m, \tbW_m$ given in~\eqref{eq:defs},
\begin{align}
\bA \bQ_m + \tbW_m = \bQ_m \tbT_m + [\tbT_m]_{m+1,m} \ \bq_{m+1}\pmb{b}_m^\dagger + \bQ_m \bE_m \label{eq:arnfact}
\end{align}
where  $\pmb{b}_m$ is the $m^{th}$ elementary vector, and $\bE_m = [\bQ_m^\dagger \tbW_m]_{U}$. We can further simplify the above, using the fact that $\bQ_m^\dagger \bQ_m = \bI_m $ and $\bQ_m^\dagger \bq_{m+1} = \bo$,
\begin{align*}
\bQ_m^\dagger \bA \bQ_m + \bQ_m^\dagger \tbW_m = \tbT_m +  \bE_m
\end{align*}
Using the definition of $\bE_m$, we write,
\begin{align*}
\bQ_m^\dagger \bA \bQ_m  &=  \tbT_m +  [\bQ_m^\dagger \tbW_m]_{U} -  \bQ_m^\dagger \tbW_m  \nonumber \\
&= \tbT_m - \tbE_m \triangleq \bC_m
\end{align*} 
where $\tbE_m = [\bQ_m^\dagger \tbW_m]_{SL} $, as defined in~\eqref{eq:defs}.  \\

\noindent $(P2):$ 
Noting that $\tbT_m + \bE_m = \bC_m + \bQ_m^\dagger \tbW_m$, we rewrite~\eqref{eq:arnfact} as, 
\begin{align*}
\bA \bQ_m - \bQ_m\bC_m =  [\tbT_m]_{m+1,m} \ \bq_{m+1}\pmb{b}_m^\dagger  - ( \bI_M -	\bQ_m \bQ_m^\dagger )\tbW_m
\end{align*}
Multiplying the latter equation by $\bs_i^{(m)}$, and using the fact that $\bC_m \bs_i^{(m)} = \lambda_i^{(m)} \bs_i^{(m)} $, and $\bQ_m \bs_i^{(m)} = \bth_i^{(m)} $
\begin{align*}
\bA \bth_i^{(m)} - & \lambda_i^{(m)} \bth_i^{(m)} \nonumber \\
 &=  [\tbT_m]_{m+1,m} \ \bq_{m+1}\pmb{b}_m^\dagger \bs_i^{(m)}  - ( \bI_M -	\bQ_m \bQ_m^\dagger )\tbW_m \bs_i^{(m)}
\end{align*}
Finally, the desired residual is upper bounded as,
\begin{align*}
\Vert \bA & \bth_i^{(m)} -  \lambda_i^{(m)} \bth_i^{(m)} \Vert_2^2 \nonumber \\
& \leq  ([\tbT_m]_{m+1,m} |\pmb{b}_m^\dagger \bs_i^{(m)}|)^2  + \Vert ( \bI_M -	\bQ_m \bQ_m^\dagger )\tbW_m \bs_i^{(m)}\Vert_F^2 \nonumber \\
& \leq  ([\tbT_m]_{m+1,m} | [\bs_i^{(m)}]_m | )^2  + \Vert  \bI_M -	\bQ_m \bQ_m^\dagger \Vert_F^2 \Vert \tbW_m \Vert_F^2 
\end{align*}
where the last inequality follows from $\Vert \bB_1 \bB_2 \bx \Vert_2^2 \leq  \Vert \bB_1 \Vert_F^2 . \Vert \bB_2 \Vert_F^2 . \Vert \bx \Vert_2^2 $ \\

\noindent $(P3):$ 
The proof immediately follows by noting that $\Vert  \bI_M -	\bQ_m \bQ_m^\dagger \Vert_F^2 \rightarrow 0 $ and $[\tbT_m]_{m+1,m} \rightarrow 0$, as $m \rightarrow M$, thereby implying that $\Vert \bA  \bth_i^{(M)} -  \lambda_i^{(M)} \bth_i^{(M)} \Vert_2^2  \ll 1 . $ 

\subsection{Proof of Lemma~\ref{lem:pert}} \label{app:2}
The proof follows from a direct application of the Bauer-Fike Theorem~\cite[Th.~7.2.2]{golub_matcomp_96}. Let $\bC_m = \bS_m \bLam_m \bS_m^{-1}$ be the diagonalizable matrix in question, and $\bT_m = \bC_m + \bP_m $ the ``perturbed'' matrix. Then, every eigenvalue $\tilde{\lambda}$ of $\bT_m$ satisfies, 
\begin{align*}
\vert \tilde{\lambda} - \lambda \vert^2 \leq \Vert \bS_m \Vert_2^2 . \Vert \bS_m^{-1} \Vert_2^2 . \Vert \bP_m \Vert_2^2 = \Vert \bQ_m^\dagger \tbW_m  \Vert_2^2 
\end{align*}
where $\lambda$ is an eigenvalue of $\bC_m$, and $\Vert \bB \Vert_2 \triangleq \sigma_{\max}(\bB) $ is the vector-induced matrix $2$-norm. The last equality follows from the fact that $\bS_m$ is unitary, as discussed in Lemma~\ref{lem:arnlem}. Using the fact that $\Vert \bB \Vert_2 \leq \Vert \bB \Vert_F $, we rewrite the last equation, 
\begin{align*}
\vert \tilde{\lambda} - \lambda \vert^2 &\leq \| \bQ_m^\dagger \tilde{\bW}_m \|_F^2  \leq \| \bQ_m \|_F^2  \| \tilde{\bW}_m \|_F^2 = m \| \tilde{\bW}_m \|_F^2
\end{align*}
This concludes the proof.

\subsection{Proof of Lemma~\ref{lem:bfsol} } \label{app:3}
Note that there is not loss in optimality by assuming the $g \in \mathbb{R}_+$. Moreover, exploiting the structure of $h_o$, the globally optimal solution can be found by optimizing for $\bff$, assuming $g$ is fixed (and vice) versa, i.e., 
\begin{align}
\bff^\star \overset{\triangle}{=}& \underset{\bff}{\argmin} \ g^2(\bff^\dagger \bff )  - 2g \Re(\bff^\dagger \tbgam_1 ), \ \st \ [\bff]_i =  1 /\sqrt{M} \ e^{j\phi_i}  \nonumber \\
\overset{(a)}{\Leftrightarrow} \  \lbrace \phi_i^\star \rbrace =& \underset{\lbrace \phi_i \rbrace}{\argmax} \ 1/\sqrt{M} \ \Re\left( \sum_{i=1}^M  r_i \ e^{j(\theta_i - \phi_i)} \right) \nonumber \\
 \lbrace \phi_i^\star \rbrace =& \underset{\lbrace \phi_i \rbrace}{\argmax } \ \sum_{i=1}^M   \Re\left(e^{j(\theta_i - \phi_i)} \right) = \lbrace \theta_i   \rbrace \nonumber 
\end{align}
where (a) follows from applying the one-to-one mapping $[\bff]_i \rightarrow 1 /\sqrt{M} \ e^{j\phi_i} , \forall i  $. Thus, $[\bff^\star]_i =  1 /\sqrt{M} \ e^{j\theta_i}, \forall i $. Plugging $\bff^\star$ into the original problem, the optimization of $g$ is a simple unconstrained quadratic problem, 
\begin{align}
g^\star \overset{\triangle}{=}& \ \underset{g}{\argmin} \ g^2 - 2g ( \Vert \tbgam_1 \Vert_1 / \sqrt{M} ) = \Vert \tbgam_1 \Vert_1 / \sqrt{M}
\end{align}


\subsection{Proof of Corrollary~\ref{cor:pert} } \label{app:5}
The proof consists of finding a closed-from expression for $\tbW_m$ as a function of $\be^{(t)}_l$ and $\be^{(r)}_l$, and applying the result of Lemma~\ref{lem:pert}. Note that $\tbw_l$ in~\eqref{eq:echo} can represent any distortion, and by comparing $\bp_l$ in both  \eqref{eq:echo} and \eqref{eq:pl}, can  infer that $\tbw_l = -\bH^\dagger \bH \be_l^{(t)} - (1/d)\bH^\dagger \be_l^{(r)}$. Thus, $\tbW_m$ in~\eqref{eq:defs} can be written as, 
\begin{align*}
\tbW_m &= - \bH^\dagger \bH [\be_1^{(t)}, \cdots, \be_m^{(t)} ] - (1/d)\bH^\dagger  [\be_1^{(r)}, \cdots, \be_m^{(r)} ] \nonumber \\
&\triangleq - \bH^\dagger \bH \bE^{(t)} - (1/d)\bH^\dagger \bE^{(r)}
\end{align*}
Then using properties of the Frobenius norm,
\begin{align}
\Vert \tbW_m \Vert_F &\leq \Vert \bH \Vert_F^2 \Vert \bE^{(t)} \Vert_F + (1/d) \Vert \bH \Vert_F  \Vert \bE^{(r)} \Vert_F \label{eq:cor0}
\end{align}
On the other hand, recall that $\be^{(t)}_l = \bq_l - \tbf_l \tilde{g}_l $ and $\be^{(r)}_l = \tbs_l - \tbw_l \tilde{u}_l $. Thus, using the results of Sec.~\ref{sec:sd1d},
\begin{align*}
\Vert \be^{(t)}_l \Vert_2 &\leq \Vert \bq_l \Vert_2 + \Vert \tbf_l \tilde{g}_l \Vert_2 \leq 2 \\
\Vert \be^{(r)}_l \Vert_2 &\leq \Vert d \bH \tbf_l \tilde{g}_l \Vert_2 + \Vert \tbw_l \tilde{u}_l \Vert_2 \leq 1 + d \Vert \bH \Vert_F 
\end{align*}
and it follows that 
\begin{align} \label{eq:cor1}
\Vert \bE^{(t)} \Vert_F \leq 2 \sqrt{m},  \ \Vert \bE^{(r)} \Vert_F \leq \sqrt{m} (1 + d \Vert \bH \Vert_F )
\end{align}
The upper bound follows by combining~\eqref{eq:cor0} and~\eqref{eq:cor1}.


\ifCLASSOPTIONcaptionsoff
  \newpage
\fi

\addcontentsline{toc}{chapter}{Bibliography}
\bibliographystyle{ieeetr}
\bibliography{ref.bib}

\end{document}